\documentclass[twoside,11pt]{article}

\usepackage{jmlr2e}

\usepackage{amsfonts} 
\usepackage{amsmath}
\usepackage{amssymb}
\usepackage{graphicx} 
\usepackage{natbib}
\usepackage{lscape}
\usepackage{color}
\usepackage{subfigure}
\usepackage{multirow}
\usepackage{booktabs}
\usepackage{dsfont}

\DeclareMathOperator{\Y}{\boldsymbol{Y}}
\DeclareMathOperator{\y}{\mathbf{y}}

\DeclareMathOperator{\bbeta}{\boldsymbol{\eta}}

\DeclareMathOperator{\f}{\mathbf{f}}

\DeclareMathOperator{\btheta}{\boldsymbol{\theta}}

\DeclareMathOperator{\balpha}{\boldsymbol{\alpha}}

\DeclareMathOperator{\x}{\mathbf{x}}

\DeclareMathOperator{\cov}{\mathrm{Cov}}

\DeclareMathOperator{\0}{\boldsymbol{0}}

\DeclareMathOperator{\diag}{\mathrm{diag}}
\DeclareMathOperator{\bell}{\boldsymbol{\ell}}

\DeclareMathOperator{\nW}{\mathrm{W}}
\DeclareMathOperator*{\argmax}{arg\,max}

\ShortHeadings{Gaussian process regression with heteroscedastic Student-{\it t} model}{Marcelo Hartmann \& Jarno Vanhatalo}
\firstpageno{1}

\begin{document}

\title{Laplace approximation and the natural gradient for Gaussian process regression with the heteroscedastic Student-{\it t} model}

\author{\name Marcelo Hartmann \email marcelo.hartmann@helsinki.fi \\
       \addr Department of Mathematics and Statistics \\
       University of Helsinki \\
       Helsinki, FI-00014, Finland
       \AND
       \name Jarno Vanhatalo \email jarno.vanhatalo@helsinki.fi \\
       \addr Department of Mathematics and Statistics \\ Department of Biosciences \\
       University of Helsinki\\
       Helsinki, FI-00014, Finland
       }

\editor{}

\maketitle

\begin{abstract}
This paper considers the Laplace method to derive approximate inference for the Gaussian process (GP) regression in the location and scale parameters of the Student-$\it{t}$ probabilistic model. This allows both mean and variance of the data to vary as a function of covariates with the attractive feature that the Student-$\it{t}$ model has been widely used as a useful tool for robustifying data analysis. The challenge in the approximate inference for the GP regression with the Student-$\it{t}$ probabilistic model, lies in the analytical intractability of the posterior distribution and the lack of concavity of the log-likelihood function. We present the natural gradient adaptation for the estimation process which primarily relies on the property that the Student-$\it{t}$ model naturally has orthogonal parametrization. 
Due to this particular property of the model, we also introduce an alternative Laplace approximation by using the Fisher information matrix. 
According to experiments this alternative approximation provides very similar posterior approximations and predictive performance when compared to the traditional Laplace approximation. We also compare both of these Laplace approximations with the Monte Carlo Markov Chain (MCMC) method. Moreover, we compare our heteroscedastic Student-$\it{t}$ model and the GP regression with the heteroscedastic Gaussian model. We also discuss how our approach can improve the inference algorithm in cases where the probabilistic model assumed for the data is not log-concave. \\[0.1cm]
\end{abstract}  


\section{Introduction}

Numerous applications in statistics and the machine learning community are fraught with datasets where some data points appear to strongly deviate from the bulk of the remaining. Usually those points are referred to outliers and in many cases the presence of outliers can drastically change the final result of data analysis \citep{atkinson:2000}. It is known that, if the probabilistic model for the data is not robust, in the sense of reducing outlier influence, inference for the probabilistic model parameters can be strongly biased and consequently prediction power is reduced \citep{finetti:1961, west:1984, atkinson:2000}. 

The Student-$\it{t}$ model \citep{stud:1908} is a three-parameter heavy-tailed probabilistic model with the outlier-prone property (robustness) in the sense of \cite{dawid:1973} and \cite{hagan:1979}. That is, the effect of a group of observations that deviates from the rest of its bulk becomes negligible as that group of observations approaches infinity. The degree of robustness of the model is directly related to the degrees-of-freedom parameter (shape parameter) $\nu$. The smaller the values of $\nu$, the more robust the model is in the presence of outliers \citep{hagan:1979, thais:2008}.

Due to the particular outlier-prone property of the Student-$\it{t}$ model,  much research has been focused on regression models (linear and non-linear) where the error term is assumed to be distributed according to the Student-$\it{t}$ model. \cite{Kennet:1989}, \cite{gew:1993} and \cite{fernandes:1999}, consider multivariate linear regression models where the error distribution is assumed to follow the Student-$\it{t}$ probabilistic model. They highlight important aspects such as goodness of fit and inferential difficulties in both Bayesian and non-Bayesian approaches. 
\cite{michael:2005} apply variational approximation to the posterior distribution of the regression parameters. \cite{thais:2008} obtain the Fisher information matrix and the Jeffrey's prior distribution \citep{jeffreys:1998} for the vector of parameters in the multivariate regression model with the Student-$\it{t}$ model. The study of \cite{wang:2016} is similar to that of \cite{thais:2008}, but they focus on the reference prior \citep{Bernardo:79} for the vector of parameters and prove that the posterior distribution for all parameters in the model is improper.

In Gaussian process (GP) regression, the Student-$\it{t}$ model has been applied with the same aforementioned principles, but instead the focus is on the treatment of the location parameter as an unknown function which follows a Gaussian process prior \citep{van:2009}. In this case, the analytical intractability of the posterior distribution with lack of concavity in the log-likelihood function brings difficulties to the estimation process.
The early works of \cite{neal:1997} consider the scale-mixture representation \citep{gew:1993} which enables more efficient MCMC methods via Gibbs sampling. \cite{van:2009} and \cite{jyla:2011} consider faster approximation methods for the posterior distribution of the Gaussian process, by either considering the Laplace method \citep{laplace, laplace2}, variational-Bayes \citep{mackay:2002, bishop:2006} or expectation-propagation (EP) \citep{minka01, minka:2001}. They point out that, since the log-likelihood function of the Student-$\it{t}$ model is not log-concave, the posterior distribution of the Gaussian process can present multimodality which makes the implementation of the Laplace method and EP more challenging than with log-concave likelihoods. The variational-Bayes approximation has a stable computational implementation but the approximation underestimates posterior variance \citep{jyla:2011}. More generally, a detailed analysis carried out by \cite{fernandes:1999} reveals that parameter inference in both Bayesian and non-Bayesian settings of multivariate regression models with Student-$\it{t}$ errors can be challeging. Firstly because the likelihood can be unbounded for small values of $\nu$ and secondly, due to the possibility of multimodality in the likelihood function with certain combinations of the parameters. 

This work is developed following the same lines of \cite{van:2009}. However we use Gaussian process priors to model both the location and the scale parameters of the Student-$\it{t}$ probabilistic model. This is an important case in which both the mean and variance of the data vary as a function of covariates with the attractive property that the Student-$\it{t}$ probabilistic model is robust. 
We focus on Laplace's method to approximate the posterior distribution of the Gaussian process and inferences are also done using it. The difficulty in the estimation process of the parameters of the Laplace approximation, discussed by  \cite{van:2009} and \cite{jyla:2011}, is circumvented by firstly noting that the location and scale parameters of the Student-$\it{t}$ model are orthogonal \citep{cox:1987, huz:1956, achcar:1994}. This particular property of the Student-$\it{t}$ model will readily allow us to propose an efficient inference algorithm for the Laplace approximation based on the natural gradient of 
\cite{amari:1998} (also known as the Fisher score algorithm in Statistics). 

In this paper, we also propose an alternative Laplace approximation for the posterior distribution of the Gaussian process model. This approximation uses the Fisher information matrix in place of the  Hessian matrix of the negative log-likelihood function. 
Moreover, the alternative Laplace approximation also suggests that the approximate marginal likelihood, which is now based on the Fisher information matrix, offers an alternative way to perform type-II maximum a posteriori (MAP) estimation for the parameters of the probabilistic model and the Gaussian process hyperparameters.

The inference algorithm for estimating the parameters of the Laplace approximation presented here is general. It closely follows the stable implementation of the Laplace approximation for log-concave likelihoods presented by \cite{Rasmussen+Williams:2006} with only minor modifications and, hence, generalizes this stable algorithm for general non-log-convace and multivariate Gaussian process models as well. These general properties are also attractive for other types of models and, hence, we present an example of orthogonal reparametrization for the Weibull probabilistic model and discuss its benefits before introducing the heteroscedastic Student-$\it{t}$ model.


The paper is organized as follows: in Section \ref{sec:parmodel} we review some definitions and examples of orthogonal parametrization for statistical models in the sense of \citet[][page 207, Section 4.31]{jeffreys:1998} and \cite{cox:1987}. This concept is needed to introduce an alternative way of improving inference in Gaussian process models. Section \ref{sec:3} presents the Student-$\it{t}$ probabilistic model and how the heteroscedastic Gaussian process regression is built. The traditional Laplace approximation with its variant based on the Fisher information matrix is presented in Section \ref{sec:4}. We also present the approximate marginal likelihood based on the Fisher information in this section. In Section \ref{sec:5}, we tackle the natural gradient adaptation for finding the parameters of both Laplace approximations. The performance of these approximations and other models are evaluated in Section \ref{sec:6}, where we examine the quality of these approximations with a simulated example and several real datasets. Section \ref{sec:7} closes the paper with the discussion and concluding remarks.


\section{Aspects of orthogonal parametrization for statistical models} \label{sec:parmodel}

This section presents the definition of orthogonal parameters and the equations to find orthogonal parametrization of a probabilistic model \citep{huz:1956, cox:1987}. These ideas will be useful later, when we identify that the Student-$\it{t}$ model directly possesses such a property. One selected example is also presented in order to illustrate and clarify concepts of reparametrization in statistical modelling. We end this section by discussing these examples and other aspects of parametric transformations. 

During the middle eighties to the end of nineties, a large amount of work in statistics focused in parameter transformation methods for statistical models \citep{cox:1987, achcar:1990, achcar:1994, kass:1994, mackay:1998}. In both Bayesian and frequentist 
inference, the performance of numerical procedures and the accuracy of approximation methods (e.g. Laplace approximation) are usually affected by the choice of the parametrization in the probabilistic model. See for example, \cite{cox:1987}, \cite{kass:1994} and \cite{mackay:1998}. In this sense, it is often highly benefitial to identify a new parametrization for a probabilistic model so that the posterior density or the likelihood function are as near as possible to a Gaussian. 


To improve the Gaussian approximation for the posterior distribution or the likelihood function, different methods have been proposed in the literature. We cite a few of them here. For instance, the orthogonal reparametrization defined by Jeffreys in 1939 \citep[][page 207, Section 4.31]{jeffreys:1998} and later investigated by \cite{huz:1950}, \cite{huz:1956} and \cite{cox:1987}, improves the "normality" of the likelihood function by choosing a new parametrization such that the Fisher information matrix is diagonal. This means that the likelihood function is better behaved in the sense that the distribution of the maximum likelihood estimators converges faster to a Gaussian density \citep{cox:1987}. An other method, as presented by \cite{achcar:1994}, proposes a reparametrization such that the Fisher information is constant. In the Bayesian context, this implies a uniform Jeffreys' prior for the parameters \citep{box:1973}.

In what follows, we assume a random variable $Y$ with a probability density function $\pi_Y(y|\balpha)$, where $\balpha = [\alpha_1, \ldots, \alpha_p]^T$ $\in$ $\mathcal{A}\subseteq \mathbb{R}^p$ is the set of real continuous parameters. We also consider that the regularity conditions hold for the probabilistic model $\pi_Y(y|\balpha)$ \citep[see][Definition 2.78, page 111]{mark:2011},
\\




\noindent
{\bf Definition 1 (Fisher information matrix)}. Given that the regularity conditions hold, the matrix $I(\balpha)$ with elements
\begin{align}
I_{i, j} (\balpha) = \mathbb{E}_{Y|\balpha} \left[-\dfrac{\partial^2\log \pi_{_Y}(Y|\boldsymbol{\alpha})}{\partial \alpha_i \partial \alpha_j} \right] 
\end{align}
is called Fisher information matrix. Note that the matrix $I(\balpha)$ is the expected value of the Hessian matrix of the negative log-density function. By definition, this matrix is symmetric and positive-definite. Its inverse is a covariance matrix which provides the Cram\'{e}r-Rao lower bound for the class of unbiased estimators \citep[see][Section 2.3 and 5.1.2 for details]{mark:2011}. \\

\noindent
{\bf Definition 2 (Orthogonal parameters}). The set of parameters $\balpha$, in the probabilistic model $\pi_Y(y|\balpha)$, are said to be orthogonal if the Fisher information matrix $I(\balpha)$ is diagonal, that is,
\begin{align}
\mathbb{E}_{Y|\balpha} \left[-\dfrac{\partial^2\log \pi_{_Y}(Y|\boldsymbol{\alpha})}{\partial \alpha_i \partial \alpha_j} \right] = 0
\end{align}
for all $i$, $j$ such that, $i \neq j$. It can also be said that the probabilistic model $\pi_Y(\cdot|\balpha)$ possesses orthogonal parametrization. 

\subsection*{Equations for finding orthogonal parameters \citep{huz:1956}}

\noindent
Consider a probabilistic model $\pi_{_Y}(y|\balpha)$ where the regularity conditions hold.
Let the new parametrization $\bbeta$ $=$ $[\eta_1, \cdots, \eta_p]^T$ $=$ $F(\balpha)$ be a bijective differentiable map (with differentiable inverse map) of $\balpha$. Rewrite the probabilistic model of $Y$ in the new parametrization as follows, 
\begin{align} \label{eq:der1}
\log \pi_{_Y}(y|\bbeta) &= \log \pi_{_Y}(y|F^{-1}(\bbeta)) \nonumber \\
 &= \log \pi_{_Y}(y|\balpha(\bbeta)).
\end{align}
The second derivatives of \eqref{eq:der1} w.r.t $\eta_i$ and $\eta_j$ leads to
\begin{align} \label{eq:der2}
\dfrac{\partial^2}{\partial \eta_i \partial \eta_j} \log \pi_{_Y}(y|\bbeta) &= \sum_{k = 1}^p \sum_{l = 1}^p \dfrac{\partial^2}{\partial \alpha_k \partial \alpha_l} \log \pi_{_Y}(y|\balpha)\dfrac{\partial \alpha_k}{\partial \eta_i}\dfrac{\partial \alpha_l}{\partial \eta_j} \nonumber \\
&\phantom{=} + \sum_{k=1}^p \dfrac{\partial}{\partial \alpha_k}\log \pi_{_Y}(y|\balpha)\dfrac{\partial^2 \alpha_k}{\partial \eta_i \partial \eta_j}
\end{align}
%
Take the expectation $\mathbb{E}_{Y|\balpha(\bbeta)}[\cdot]$ with the negative sign in both sides of equation \eqref{eq:der2}. Given that the regularity conditions hold, we have
\begin{align}
I_{i, j}(\bbeta) = \sum_{k = 1}^p \sum_{l = 1}^p \dfrac{\partial \alpha_k}{\partial \eta_i}\dfrac{\partial \alpha_l}{\partial \eta_j} I_{k, l}(\balpha).
\end{align}
If we want the parameters $\bbeta$ to be orthogonal we set,
\begin{align} \label{eq:orteq}
I_{i, j}(\bbeta) &= 0.
\end{align}
for $i \neq j$. In order to find such parametrization we need to solve the system of $p(p-1)/2$ first order partial differential equations, with the $\alpha_i$, $i$ $=$ $1,\ldots,p$ as dependent variables. \\

\noindent
{\bf Example}. We present an example of orthogonal parametrization with the Weibull model. Then, we compare the Laplace approximation of the posterior densities in the common parametrization and in the orthogonal parametrization. Let $Y|\alpha_1, \alpha_2$ $\sim$ $\mathcal{W}(\alpha_1, \alpha_2)$ denote a random variable following the Weibull distribution with common parametrization $\alpha_1$ and $\alpha_2$. Then the probability density function of $Y$ is given by,
\begin{align}
\pi_{_Y}(y|\alpha_1, \alpha_2) = \alpha_1\alpha_2(\alpha_2 y)^{\alpha_1 - 1}\exp(-(\alpha_2 y) ^{\alpha_1}) \mathds{1} _{(0, \infty)}(y)
\end{align}
for $\alpha_1, \alpha_2 \in (0, \infty)$. The Fisher information matrix for this model was obtained by \cite{gupta:2006} (in their notation, $\alpha_1 = \beta$ and $\alpha_2 = \theta$, see page 3131). We now consider that, in the new parametrization $[\eta_1, \eta_2]^T$ $=$ $F(\alpha_1, \alpha_2)$ the Fisher information matrix is diagonal.
To find this new parametrization, we start with equation \eqref{eq:orteq}, which gives
\begin{align} \label{eq:exG}
0 &= I_{1, 2}(\eta_1, \eta_2) \nonumber \\
&= \dfrac{\partial \alpha_1}{\partial \eta_1}\dfrac{\partial \alpha_1}{\partial \eta_2} I_{1, 1}(\balpha) + 
\dfrac{\partial \alpha_1}{\partial \eta_1}\dfrac{\partial \alpha_2}{\partial \eta_2} I_{1, 2}(\balpha) + 
\dfrac{\partial \alpha_2}{\partial \eta_1}\dfrac{\partial \alpha_1}{\partial \eta_2} I_{2, 1}(\balpha) + 
\dfrac{\partial \alpha_2}{\partial \eta_1}\dfrac{\partial \alpha_2}{\partial \eta_2} I_{2, 2}(\balpha).
\end{align}
Now, we fix $\alpha_1 = h_1(\eta_1)$ and choose $\alpha_2 = h_2(\eta_1, \eta_2)$, such that $\eta_1$ and $\eta_2$ are orthogonal parameters 
(we also could fix $\alpha_2 = h_2(\eta_2)$ and choose $\alpha_1 = h_1(\eta_1, \eta_2)$, such that $\eta_1$ and $\eta_2$ are orthogonal parameters).
We choose $\alpha_1$ = $\exp(\eta_1)$. Thus, given the elements of the Fisher information matrix in \cite{gupta:2006} (page 3134), equation \eqref{eq:exG} becomes,
\begin{align}
0 &= \exp(\eta_1) I_{1, 2}(\balpha) + \dfrac{\partial \alpha_2}{\partial \eta_1} I_{2, 2}(\balpha) \Longleftrightarrow\nonumber \\
& c \hspace{0.05cm} \partial \eta_1 \exp(-\eta_1) = - \hspace{0.05cm} \partial \alpha_2/\alpha_2
\end{align}
whose solution is
\begin{align} \label{eq:solex2}
c \hspace{0.05cm} \exp(-\eta_1) + c \hspace{0.05cm} z(\eta_2) &= \ln \alpha_2.
\end{align}
where $c = 1 + \psi(1)$ and $\psi(\cdot)$ is the digamma function. $z(\eta_2)$ is our integration constant and we set $z(\eta_2) = \eta_2$. Rearrange equation \eqref{eq:solex2} to get
\begin{align}
\alpha_2 = \exp\big(c \hspace{0.02cm} \exp(-\eta_1) + c \hspace{0.06cm} \eta_2 \big).
\end{align}
Hence the Weibull model with orthogonal parameters is given by,
\begin{align}
\pi_{_Y}(y|\eta_1, \eta_2) &= \exp(\eta_1 + c \hspace{0.05cm} e^{-\eta_1} + c \hspace{0.05cm} \eta_2)(\exp(c \hspace{0.05cm} e^{-\eta_1} + c \hspace{0.05cm} \eta_2) y)^{\exp(\eta_1) - 1} \nonumber \\
&\phantom{=} \times \exp\big(-(\exp(c \hspace{0.05cm} e^{-\eta_1} + c \hspace{0.05cm} \eta_2)y)^{\exp(\eta_1)}\big) \mathds{1}_{(0, \infty)}(y).
\end{align}
The parametrization $(\eta_1, \eta_2)$ is now unconstrained (on $\mathbb{R}^2$) with diagonal Fisher information matrix and the transformation $[\eta_1,  \eta_2]^T = [\log \alpha_1, (\log \alpha_2)/c - 1/\alpha_1]^T$ is one-to-one.
\begin{center}
\begin{figure}[!htb]
\setlength{\parindent}{0.1cm}
\includegraphics[scale = 0.50]{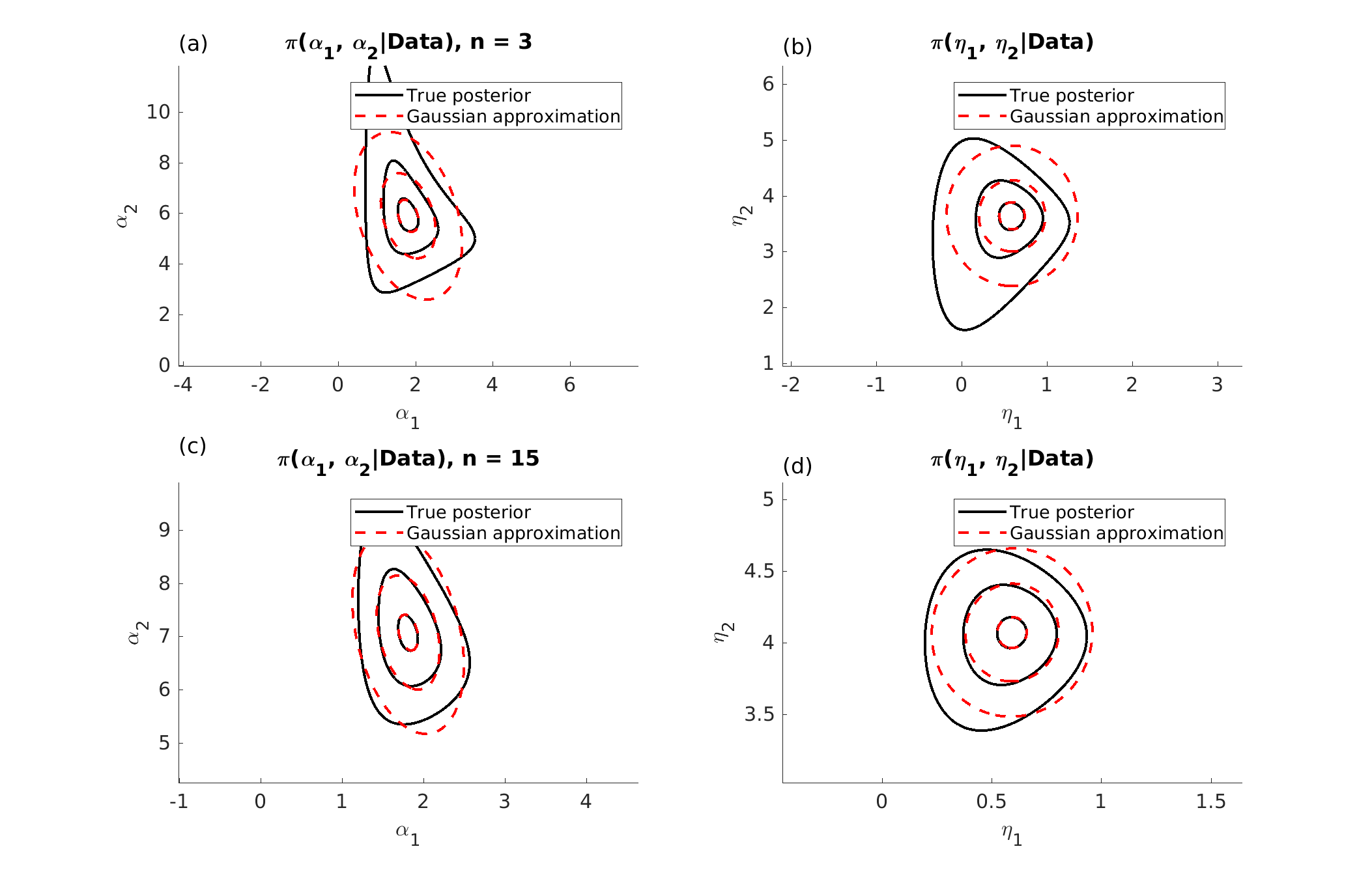}
\caption{Panel (a) and (b) shows the approximate posterior distribution of $\alpha_1, \alpha_2|\y$ and $\eta_1, \eta_2|\y$ using Laplace's method with $n = 3$. In panels (c) and (d) we redo the same but with the larger sample size $n = 15$. For the sample size $n = 15$, both approximations are much closer to a Gaussian as seen in panel (c) and (d). However, in panel (d), the approximate Gaussian is still close to being independent, which does not happen in the approximate posterior for the parametrization $(\alpha_1, \alpha_2)$ in panel (c).}
\label{fig:fig_orth}
\end{figure}
\end{center}

In order to compare the Laplace approximation of the posterior densities in the two parametrizations $(\alpha_1, \alpha_2)$ and $(\eta_1, \eta_2)$, we simulated data $Y_i$ $\sim$ $\mathcal{W}$ $(7, 1.5)$ with two different sample sizes, $n = 3$ and $n = 15$. Figure \ref{fig:fig_orth} displays the comparisons between the approximate posterior distribution of $\alpha_1, \alpha_2|\y$ and $\eta_1, \eta_2|\y$, using Laplace's method with sample sizes $n = 3$ and $n = 15$. We note that, the shape of the posterior distribution in parametrization $(\eta_1, \eta_2)$ is visually closer to an independent Gaussian density than the shape of the posterior distribution in parametrization $(\alpha_1, \alpha_2)$.

In the example presented above, the new parametrization of the statistical model improved the Laplace approximation. As pointed out by \cite{mackay:1998}, the effect of the reparametrization in probabilistic models can also lead to better approximation for the marginal likelihood. If the posterior density is well approximated by a Gaussian, then the Laplace approximation for the marginal likelihood is also better. In real-world scenarios, where complex models impose challenges, it would be beneficial to search for a parametrization of the probabilistic model so that approximation methods and numerical procedures are improved. Hence, the necessity to engineer a complex inference algorithm could be alleviated whereas existing methods could be ameliorated.

\section{Gaussian process regression with the heteroscedastic Student-$t$ model} \label{sec:3}

In this section, we highlight the basic properties of the Student-$\it{t}$ probabilistic model, which are useful to clarify how model building with Gaussian process priors is done. We present how the Student-$\it{t}$ model is parametrized and where the Gaussian process prior is introduced in the parameters of the probabilistic model to build the Gaussian process regression. 

\subsection{Student-$\it{t}$ model and basic properties}

Let us denote by $Y|\mu, \sigma, \nu$ $\sim$ $\mathcal{S}(\mu, \sigma, \nu)$ a random variable which follows the Student-$\it{t}$ probabilistic model with the
location ($\mu$), scale ($\sigma$) and degrees-of-freedom ($\nu$) parameters.
Then the probability density function of $Y|\mu, \sigma, \nu$ is given by,
\begin{equation}\label{eq:stmodel_a}
\pi(y|\mu, \sigma, \nu) = \frac{\Gamma\left(\tfrac{\nu+1}{2}\right)}{\Gamma(\frac{\nu}{2}) \sigma\sqrt{\pi \nu}}\left[1 + \frac{1}{\nu}\left(\frac{y-\mu}{\sigma}\right)^2 \right]^{-\frac{\nu+1}{2}}
\end{equation}
for $\mu$ $\in$ $\mathbb{R}$, $\sigma$ $>$ $0$ and $\nu$ $>$ $0$. The expected value of $Y$ is $\mathbb{E}(Y) = \mu$ which exists only for $\nu > 1$ (otherwise $Y$ is non-integrable). The variance of $Y$ is $\mathbb{V}(Y) = \sigma^2 \nu/(\nu - 2)$, which only depends on the scale and degrees-of-freedom paramaters. If $\nu \leqslant 2$ then $\mathbb{V}(Y) = \infty$. The degrees-of-freedom parameter controls the "thickness" of the tails in the probability density function \eqref{eq:stmodel_a}. The smaller the values of $\nu$, the more robust the Student-$\it{t}$ model is in presence of outliers \citep{ohagan78}. If $\nu \rightarrow \infty$ the model \eqref{eq:stmodel_a} converges to a Gaussian density function with parameters $(\mu, \sigma^2)$ \citep{thais:2008}.

The Fisher information matrix for the set of parameters $(\mu, \sigma, \nu)$ was obtained by \cite{thais:2008} (page 332, Proof of Theorem 2) and we note that entries $(1, 2)$ and $(1, 3)$ of the Fisher information matrix are zero. Therefore, this means that $(\mu, \sigma)$ and $(\mu, \nu)$ are pairs of orthogonal parameters. Although the model does not possesses full orthogonality, since entry $(2, 3)$ of the Fisher information matrix
is non-zero, this particular property of the model will be useful later in Section 5. In that section, we tackle a computational implementation to efficiently perform approximate inference with this model.
%
%

\subsection{Gaussian process regression in the location and scale parameter}

Consider the regression model for a set of data $\Y^T = [Y_1 \ \cdots \ Y_n] \in \mathbb{R}^n$ that satisfies
\begin{equation} \label{eq:reg}
Y_i = f_1(\x_i) + \varepsilon_i\exp(f_2(\x_i))
\end{equation}
for $i = 1$, $\hdots$, $n$ where $n$ is the number of observations and $\x_i$ is the $i^{th}$ vector of covariates. Assume that $f_1(\cdot)$ and $f_2(\cdot)$ follow independent zero-mean Gaussian process priors. This implies that $\f_1^T$ $=$ $[f_1(\x_1) \cdots f_1(\x_n)]$ $\sim$ $\mathcal{N}(\0, K_1)$ and $\f_2^T$ $=$ $[f_2(\x_1) \cdots f_2(\x_n)]$ $\sim$ $\mathcal{N}(\0, K_2)$. The matrix $\lbrace K_1 \rbrace_{i, j}$ $=$ $\cov(f_1(\x_i), f_1(\x_j)|\gamma_1)$ is the covariance matrix of the process $f_1$, which depends on a vector of hyperparameters $\gamma_1$ and the matrix $\lbrace K_2 \rbrace_{i, j}$ $=$ $\cov(f_2(\x_i), f_2(\x_j)|\gamma_2)$ is the covariance matrix for the process $f_2$, which depends on a vector of hyperparameters $\gamma_2$. Now, let $\varepsilon_i|\nu$ $\stackrel{i.i.d}{\sim}$ $\mathcal{S}(0, 1, \nu)$.
%
%
Therefore for each $i$, the random variable $Y_i|f_1(\x_i), f_2(\x_i), \nu \sim \mathcal{S}(f_1(\x_i), \exp(f_2(\x_i)), \nu)$ has density function given by
\begin{equation}\label{eq:stmodel}
\pi(y_i|f_1(\x_i), f_2(\x_i), \nu) = \frac{\Gamma\left(\tfrac{\nu+1}{2}\right)}{\Gamma(\frac{\nu}{2}) \exp(f_2(\x_i))\sqrt{\pi \nu}}\left[1 + \frac{1}{\nu}\left(\frac{y_i-f_1(\x_i)}{\exp(f_2(\x_i))}\right)^2 \right]^{-\frac{\nu+1}{2}}
\end{equation}
Lets denote by $\y^T$ $=$ $[y_1$ $\cdots$ $y_n]$ the set of measured data, $\f^T$ $=$ $[\f^T_1  \f^T_2]$ the vector of all the latent function values and $\btheta^T$ $=$ $[\nu \ \gamma_1 \ \gamma_2]$ the collection of all probabilistic models parameters and covariance functions hyperparameters. Then, by the Bayes' rule, the conditional posterior distribution for $\f|\y, \btheta$ is obtained as
\begin{equation}\label{eq:post}
\pi(\f|\y, \btheta) = \tfrac{1}{\pi(\y\hspace{-0.05cm}|\hspace{-0.05cm}\btheta)} L(\y|\f, \nu )\mathcal{N}(\f_1|\0, K_1)\mathcal{N}(\f_2|\0, K_2)
\end{equation}
where
\begin{align} \label{eq:likfun}
L(\y|\f, \nu ) = \prod_{i=1}^n \pi(y_i|f_1(\x_i), f_2(\x_i), \nu)
\end{align}
is the likelihood function of $\f$ and
\begin{equation} \label{eq:margl}
\pi(\y\hspace{-0.05cm}|\hspace{-0.05cm}\btheta) = \int\limits_{\mathbb{R}^N} L(\y|\f, \nu )\mathcal{N}(\f_1|\0, K_1)\mathcal{N}(\f_2|\0, K_2) \hspace{0.05cm} \mathrm{d} \hspace{-0.03cm}\f
\end{equation}
is the marginal likelihood (the normalizing constant). Note that, expression \eqref{eq:margl} can not be solved analytically and, moreover, posterior expectations and posterior variances are not found in closed-form. Furthermore, the posterior distribution \eqref{eq:post} has dimension two times greater than the number of data points ($N = 2n$), which additionally imposes more difficulty in the implementation of any inference algorithm.

\section{Approximate inference with the Laplace method} \label{sec:4}

In this section, we present the Laplace method to perform approximate inference. This method is a useful technique for integrals arising in Bayesian inference \citep{laplace, laplace2, rue, migon:2014}. The approximation is analytical and utilizes the Gaussian density function for the approximation. The Gaussian density has desirable analytical properties such as, closed under marginalization and conditioning \citep{seber:2003, seber:2012}. In what follows, we carry out the Laplace approximation for \eqref{eq:post} and \eqref{eq:margl} using a similar approach and notation as in \cite{Rasmussen+Williams:2006}. We also present the Laplace approximation where the Hessian matrix of the negative log-likelihood function is replaced by its expected value, that is, the Fisher information matrix.

\subsection{Laplace approximation}

The Laplace approximation is based on the second-order Taylor expansion of $\log \pi(\f|\y, \btheta)$ around the mode (maximum a posteriori estimate) $\hat{\f}$ $=$ $\argmax_{\f \in \mathbb{R}^N}$ $\log$ $\pi(\f|\y, \btheta)$. 
The method yields a multivariate Gaussian approximation for the conditional posterior distribution \eqref{eq:post} given by
\begin{align} \label{eq:postap} 
\pi_{1}(\f|\y, \btheta) = \mathcal{N}\big(\hspace{-0.04cm}\f|\hat{\f}, (K^{-1} + \nW)^{-1} \big)
\end{align}
The covariance matrix $K$ of the Gaussian process priors is a block diagonal matrix whose blocks are $K_1$ and $K_2$, that is, $K$ $=$ $\diag(K_1, K_2)$. 
The matrix $\nW $ $=$ $- \nabla \nabla \log L(\y|\hat{\f}, \btheta)|_{\f = \hat{\f}}$ is the Hessian matrix of the negative log-likelihood function with respect to $\f$, which is evaluated at $\hat{\f}$. More specifically, $\nW$ is a two-diagonal banded matrix whose elements are given in Appendix \ref{app:thirdd}.

\subsection{Laplace-Fisher approximation}

Since the Student-$\it{t}$ model is regular and possesses orthogonal parametrization with respect to $\mu$ and $\sigma$, 
we follow \cite{jeffreys:1998} and \citet{rober:1992} to replace $\nW$ by its expected value $\mathbb{E}_{\Y|\f, \btheta}[\nW]$ (the Fisher information matrix) in the traditional Laplace approximation \eqref{eq:postap}. Due to the real-valued random variable $f_2(\x_i)$ in \eqref{eq:stmodel}, we have to obtain the Fisher information matrix with respect to this specific real-line parametrization. The elements of $\mathbb{E}_{\Y|\f, \btheta}[\nW]$, in this specific parametrization, are given in Appendix \ref{app:thirdd}. 
The Laplace approximation for the conditional posterior distribution \eqref{eq:post} is now given by,
\begin{align} \label{eq:postap2} 
\pi_{2}(\f|\y, \btheta) = \mathcal{N}\big(\hspace{-0.04cm}\f|\hat{\f}, (K^{-1} + \mathbb{E}_{\Y|\hat{\f}, \btheta}[\nW])^{-1}\big)
\end{align}

In case of the Laplace-Fisher approximation \eqref{eq:postap2}, $\mathbb{E}_{\Y|\hat{\f}, \btheta}[\nW]$ is diagonal with positive-elements, thus the covariance matrix $(K^{-1} + \mathbb{E}_{\Y|\hat{\f}, \btheta}[\nW])^{-1}$ is such that its diagonal elements are always smaller than the diagonal elements of $K$ (element-wise). Hence, the possible effect of larger posterior variance of the latent function values with respect to its prior variance, in the approximation \eqref{eq:postap}, vanishes (see \cite{van:2009} Section 3.4 and \cite{jyla:2011} Section 5, for details). \cite{Kass+Raftery:1995} and \cite{raftery:1996} also point out that the approximation \eqref{eq:postap2} is less precise than the approximation \eqref{eq:postap}, but it will remain accurate enough for many practical purposes. 

\subsection{Prediction of future outcomes with the Laplace approximation}
Let $Y_*|\btheta, \y$ be the value of a future outcome under the presence of covariates $\x_*$ given the data and the set of parameter $\btheta$. 
If we use the approximation \eqref{eq:postap} for \eqref{eq:post}, the approximate posterior predictive distribution of the vector of latent function values at the new point $\x_*$ is given by \citep{Rasmussen+Williams:2006}
\begin{equation} \label{eq:postpred}
\begin{bmatrix}
f_1(\x_*) \\
f_2(\x_*)
\end{bmatrix} \Big| \btheta, \y \sim \mathcal{N}\left(
\begin{bmatrix}
\mu_1(\x_*) \\
\mu_2(\x_*)
\end{bmatrix},
\begin{bmatrix}
\sigma^2_1(\x_*) & \sigma_{12}(\x_*) \\ 
\sigma_{21}(\x_*) & \sigma^2_2(\x_*)
\end{bmatrix} 
\right)
\end{equation} 
with 
\begin{equation} \label{eq:predEf}
\begin{bmatrix}
\mu_1(\x_*) \\
\mu_2(\x_*)
\end{bmatrix} = \mathbf{k}(\x_*) \begin{bmatrix}
\nabla_{\f_1} \log L(\y|\hat{\f}, \nu) \\
\nabla_{\f_2} \log L(\y|\hat{\f}, \nu)
\end{bmatrix} 
\end{equation}
and
\begin{equation} \label{eq:predVf}
\begin{bmatrix}
\sigma^2_1(\x_*) & \sigma_{12}(\x_*) \\ 
\sigma_{21}(\x_*) & \sigma^2_2(\x_*)
\end{bmatrix}
 =   \diag(k_1(\x_*), k_2(\x_*)) - 
\mathbf{k}(\x_*)
(K^{-1} + \nW)^{-1}\mathbf{k}(\x_*)^T
\end{equation}
where
\begin{equation}
\mathbf{k}(\x_*) = \begin{bmatrix}
\mathbf{k}_1(\x_*) & \0_{1, n}\\
\0_{1, n} & \mathbf{k}_2(\x_*)
\end{bmatrix}.
\end{equation}
$k_1(\x_*)$ and $k_2(\x_*)$ denote the respective variances of the latent functions $f_1(\x_*)$ and $f_2(\x_*)$ obtained from the covariance functions $\cov(f_1(\x_*)$, $f_1(\x_*)|\gamma_1)$ and $\cov(f_2(\x_*)$, $f_2(\x_*)|\gamma_2)$ respectively. $\mathbf{k}_1(\x_*)$ and $\mathbf{k}_2(\x_*)$ are $1$ by $n$ row-vectors which contain the covariances $\cov(f_1(\x_*)$, $f_1(\x_i)|\gamma_1)$ and $\cov(f_2(\x_*)$,$ f_2(\x_i)|\gamma_2)$ for $i$ $=$ $1,\hdots,n$, respectively. If we use approximation \eqref{eq:postap2} instead of \eqref{eq:postap} to approximate the posterior density \eqref{eq:post}, the approximate posterior predictive distribution \eqref{eq:postpred} has diagonal covariance matrix \eqref{eq:predVf} ($\sigma_{12}(\x_*) = \sigma_{21}(\x_*) = 0$), since $\mathbb{E}_{\Y|\hat{\f}, \btheta}[\nW]$ is diagonal. Its mean vector will be equal to \eqref{eq:predEf}, given that the mode $\hat{\f}$ remains unchanged for the same $\btheta$. 

Now, the unconditional expectation (for $\nu > 1$) and unconditional variance (for $\nu > 2$) of the future outcome at $\x_*$ are obtained as, 
\begin{align} \label{eq:predE}
\mathbb{E}(Y_*|\btheta, \y) &= \mathbb{E}[\mathbb{E}(Y_*|f_1(\x_*), f_2(\x_*), \btheta, \y)] \nonumber \\
&= \mathbb{E}[f_1(\x_*)|\btheta, \y] = \mu_1(\x_*)
\end{align}
and 
\begin{align} \label{eq:predV}
\mathbb{V}(Y_*|\btheta, \y) &= \mathbb{V}[\mathbb{E}(Y_*|f_1(\x_*), f_2(\x_*), \btheta, \y)] + \mathbb{E}[\mathbb{V}(Y_*|f_1(\x_*), f_2(\x_*), \btheta, \y)] \nonumber \\
&= \mathbb{V}(f_1(\x_*)|\btheta, \y) +\mathbb{E}\left(\dfrac{\nu}{\nu-2}\big(e^{f_2(\x_*)}\big)^2 \bigg|\btheta, \y\right) \nonumber \\ 
&= \sigma^2_1(\x_*) + \dfrac{\nu}{\nu-2} e^{2\mu_2(\x_*) + 2 \sigma^2_2(\x_*)}.
\end{align}
%


\section{On the computational implementation} \label{sec:5}

The main difficulty to make the approximation \eqref{eq:postap} and \eqref{eq:postap2} useful in practice is in the determination of $\hat{\f}$ for a given $\btheta$ (henceforth we refer to it only as $\hat{\f}$). As pointed out by \cite{van:2009} and \cite{jyla:2011}, the Student-$\it{t}$ model is not log-concave and will lead to numerical instability of classical gradient-based algorithms for finding the $\hat{\f}$ if the problem is not approached properly. Besides, the computational algorithm proposed in \cite{Rasmussen+Williams:2006} based on Newton's method relies on $\nW$ being non-negative with log-concave likelihoods. With the Student-$\it{t}$ model, the log-likelihood is not concave and Newton's method to find the maximum a posteriori $\hat{\f}$ is essentially uncontrolled and not guaranteed to converge \citep{van:2009}.
In the next subsections, we deal with the problem of finding the maximum a posteriori $\hat{\f}$ and how to choose $\btheta$ in the approximations \eqref{eq:postap} and \eqref{eq:postap2}.
 


\subsection{Natural gradient for finding the mode}

The problem of finding $\hat{\f}$ is approached by using a variant of standard gradient-based optimization methods called natural gradient adaptation \citep{amari:1998}. The method uses the curved geometry of the parametric space defined by the Riemannian metric \citep{amari:2007} which has been shown to improve efficiency and convergence of the computational algorithms \citep{amari:1998, honkela+raiko:2010}. As shown by \cite{amari:1998} and \cite{yannetal:2017}, the steepest ascent direction of a smooth function, say $h:$ $\mathcal{M}$ $\subseteq$ $\mathbb{R}^d$ $\rightarrow$ $\mathbb{R}$ in a Riemannian manifold $(\mathcal{M} , g)$ where $g$ is the Riemannian metric, is given by the natural gradient defined as 
%
\begin{equation} \label{eq:stepasc}
\nabla^G h(p) = G^{-1}(p) \nabla h(p).
\end{equation}
where $\nabla$ is the gradient operator and $G(\cdot)$ is the matrix of metric coefficients (positive-definite matrix $\forall p \in \mathcal{M}$). The evident challenge at this point is how to specify $G(\cdot)$, which still requires specific knowledge of the problem in question. However, it turns out that, in any regular statistical model \citep{mark:2011}, a Riemannian manifold can be obtained when the parametric space of the probabilistic model is endowed with the Fisher information matrix \citep{rao:1945, atk:1981, benmark11, calder12}. That is, the covariance between the elements of the score vector of the probabilistic model \citep{mark:2011}. Similar ideas have been successfully applied in many optimization techniques and MCMC methods. See for example works by \cite{jeenrich:1976}, \cite{amari:1998}, \cite{honkela+raiko:2010}, \cite{benmark11}, \cite{calder12}, \cite{yannetal:2017} and \cite{leon:2017}.

Now, the iterative procedure to find $\hat{\f}$ via the natural gradient is given by \citep{amari:1998, polak:2006}
\begin{align} \label{eq:upR}
\f^{\mathrm{new}} = \f + G(\f)^{-1}[\nabla \log L(\y|\f, \nu) - K^{-1} \f]
\end{align}
where $G$ is the matrix of metric coefficients. At this point, note that equation \eqref{eq:upR} is very similar to the Newton-Raphson updating scheme \cite[see][equation (3.18)]{Rasmussen+Williams:2006}. 
\begin{align} \label{eq:upW}
\f^{\mathrm{new}} &= \f - \big(\nabla \nabla_{\f} \log \pi(\f|\y, \btheta)\big)^{-1}(\nabla \log L(\y|\f, \nu) - K^{-1} \f) \nonumber \\
&= \f +(K^{-1} + \nW)^{-1}(\nabla \log L(\y|\f, \nu) - K^{-1} \f)
\end{align}
More specifically, in the case of \eqref{eq:upR}, $G$ is, by construction, always positive-definite \citep{amari:2007, rao:1945, mark:2011}, while $(K^{-1} + \nW)$ in \eqref{eq:upW} may not be, since $\nW$ is not positive-definite in the domain of the negative log-likelihood function of the Student-{\it t} model. Now, $G$ has not been specified yet and as we adopt a Bayesian approach, we would like to consider the geometry of the posterior distribution which includes the information in the likelihood and in the prior distribution. A possible Riemmanian metric with prior knowledge was used by \citet{benmark11, calder12} (page 87, Section 4.1.4, equation 4.2) and for our settings, their matrix $G(\f)$ is given by
%
\begin{align} \label{eq:metric}
G(\f) &= \mathbb{E}_{\Y|\f, \btheta}\big[-\nabla \nabla_{\f} \log \pi(\Y, \f| \btheta)\big] \nonumber \\
 &= \mathbb{E}_{\Y|\f, \btheta}\big[-\nabla \nabla_{\f} \log L(\Y|\f, \nu)\big] + \mathbb{E}_{\Y|\f,\btheta}\big[-\nabla \nabla_{\f} \log \mathcal{N}(\f_1|\0, K_1)\mathcal{N}(\f_2|\0, K_2)\big]  \nonumber \\
&= \mathbb{E}_{\Y|\f, \btheta}[\nW] + \mathbb{E}_{\Y|\f,\btheta}[K^{-1}].
\end{align}
%
Note again that, $\mathbb{E}_{\Y|\f, \btheta}[\nW]$ is the expected value of $\nW$, that is, the Fisher information matrix which has been already obtained in Section \ref{sec:4}. The second term 
$\mathbb{E}_{\Y|\f,\btheta}[K^{-1}]= K^{-1}$ is the inverse of the block diagonal covariance matrix of the Gaussian process prior. Hence, equation \eqref{eq:metric} simplifies to $G(\f) = \mathbb{E}_{\Y|\f, \btheta}[\nW]$ $+$ $K^{-1}$. Plug $G(\f)$ into equation \eqref{eq:upR} and rearrange to get
\begin{align} \label{eq:amup}
\f^{\mathrm{new}} = (K^{-1} + \mathbb{E}_{\Y|\f, \btheta}[\nW])^{-1}\big(\mathbb{E}_{\Y|\f, \btheta}[\nW]\f + \nabla \log L(\y|\f, \nu)\big)
\end{align}
which has the same structural properties as the Newton-update in \citet[equation 3.18]{Rasmussen+Williams:2006} for the binary Gaussian process classification case. Moreover, since $\mathbb{E}_{\Y|\f, \btheta}[\nW]$ is diagonal, the stable formulation of the computation algorithm provided in \cite{Rasmussen+Williams:2006} to find $\hat{\f}$ is straightforwardly applied by replacing $\nW$ with its expected value, that is $\mathbb{E}_{\Y|\f, \btheta}[\nW]$ \citep[see][Section 3.4.3, page 45]{Rasmussen+Williams:2006}. Besides, the computational cost to calculate the inverse of $(K^{-1} + \mathbb{E}_{\Y|\f, \btheta}[\nW])$ is $2\mathcal{O}(n^3)$ instead of $8\mathcal{O}(n^3)$ with the Newton-update \eqref{eq:upW}.

In case of the Gaussian process regression with the homocedastic Student-$\it{t}$ model ($f_2(\x)$ is constant), the GPML \citep{Rasmussen:2010} and GPstuff \citep{Vanhatalo:2013} software packages use the stabilized Newton algorithm to find $\hat{\f}$. In this approach the Newton direction $\mathbf{d}$ $=$ $\big(K^{-1}$ $+$ $\max(\0, \diag(\nW))\big)^{-1}$ $\nabla \log \pi(\f| \y, \btheta)$ is used \citep[see][page 3231, Section 3.2]{jyla:2011}. We see that the natural gradient adaptation uses $\mathbb{E}_{\Y|\f, \btheta}[\nW]$ in place of $\max(\0, \diag(\nW))$.

\subsection{Approximate marginal likelihood and parameter adaptation}

Note that, in equation \eqref{eq:post}, the set of parameters $\btheta$ is fixed but unknown. \cite{Rasmussen+Williams:2006} proposes a value for $\btheta$ such that $\log \pi(\y|\btheta)$ \eqref{eq:margl} is maximized. \cite{mark:1997} and \cite{van:2009} considers that, even though $\btheta$ is fixed, it is treated as an unknown quantity and so prior distributions are chosen for all its components. Our choice follows the latter and we use the maximum a posterior estimate (MAP) of $\btheta|\y$ to choose $\btheta$, that is
\begin{equation} \label{eq:th}
\hat{\btheta} = \underset{\btheta \in \Theta}{\argmax}\log \pi(\y|\btheta) + \log \pi(\btheta)
\end{equation}
where $\Theta$ is a parametric space and $\pi(\btheta)$ is the prior distribution
for $\btheta$. A closed-form expression for \eqref{eq:margl} is not known when the likelihood takes its form from the Student-$\it{t}$ model. For this reason we use Laplace's method to also approximate the marginal likelihood \eqref{eq:margl} \citep{Rasmussen+Williams:2006, rue, van:2009}. The logarithm of the marginal likelihood \eqref{eq:margl} is then approximated as 
\begin{align} \label{eq:marglap}
q_1(\y|\btheta) &= \log L(\y|\hat{\f}, \nu) - \tfrac{1}{2}\hat{\f}^TK^{-1}\hat{\f} - \tfrac{1}{2}\log |I_N + \nW \hspace{-0.1cm} K|.
\end{align}
However, since $\nW$ is not guaranteed to be positive-definite, direct evaluation of the approximate log marginal likelihood can be numerically unstable due to the last term in \eqref{eq:marglap} (see \citealt{van:2009} Section 4.2 and \citealt{jyla:2011} Section 5.4 for more details). 

Similary, as a byproduct of the approximation \eqref{eq:postap2}, the approximate log marginal likelihood in the case of the Laplace-Fisher approximation is given by 
\begin{align} \label{eq:marglapF}
q_2(\y|\btheta)  &= \log L(\y|\hat{\f}, \nu) - \tfrac{1}{2}\hat{\f}^TK^{-1}\hat{\f} - \tfrac{1}{2}\log |I_N + (\mathbb{E}_{\Y|\hat{\f}, \btheta}[\nW])^{\frac{1}{2}} K (\mathbb{E}_{\Y|\hat{\f}, \btheta}[\nW])^{\frac{1}{2}}|
\end{align}
where the last term in \eqref{eq:marglapF} is now stable to compute since $\mathbb{E}_{\Y|\f, \btheta}[\nW]$ is positive-definite. 
The formulation of the approximate log marginal likelihood \eqref{eq:marglapF} is the same as the one presented in \cite{Rasmussen+Williams:2006} (see equation 3.32, page 48), which makes its use more attractive due to its stable computational implementational. Besides, in equations \eqref{eq:marglap} and \eqref{eq:marglapF}, $\hat{\f}$ depends on $\btheta$, and the matrices $\nW$ and $\mathbb{E}_{\Y|\hat{\f}, \btheta}[\nW]$, depends on $\btheta$ and on $\btheta$ through $\hat{\f}$. \cite{Rasmussen+Williams:2006} present closed-form derivatives of \eqref{eq:marglap} w.r.t $\btheta$, which can as well be applied in the case of \eqref{eq:marglapF}. Hence, their stable computational implementation is fully applicable to the case where we set $\btheta$ by maximizing the approximate log marginal likelihood \eqref{eq:marglapF} \citep[see][Section 5.5.1, page 125]{Rasmussen+Williams:2006}. 

In Appendix \ref{app:thirdd}, we present the derivatives of $\log L(\y|\f, \nu)$ and $\nW$ w.r.t $\f_1$, $\f_2$ and $\nu$, which are needed for the computational algorithm. The derivatives of $\mathbb{E}_{\Y|\f, \btheta}[\nW]$ w.r.t $\f_1$, $\f_2$ and $\nu$ are not given since they are simple to calculate.

\section{Experiments} \label{sec:6}


This section illustrates pratical applications of the Laplace approximation \eqref{eq:postap} and the Laplace-Fisher approximation \eqref{eq:postap2} for the GP regression with the heteroscedastic Student-$\it{t}$ model. We present a simulated example to pinpoint practical differences whether conducting data analysis with the traditional Laplace approximation or with the Laplace-Fisher approximation. The predictive performance of both Laplace approximations are compared with several datasets presented in the literature. These comparisons also include the gold standard MCMC method. In the MCMC approximation, the posterior samples of \eqref{eq:post} are obtained via the elliptical slice sampler method proposed by \cite{murray:2010}. Moreover, the predictive comparisons also include the GP regression with the homoscedastic Student-$\it{t}$ model \citep{van:2009} and the GP regression with the heteroscedastic Gaussian model ($\nu \rightarrow \infty$). 

The choice of prior distributions for the Gaussian process hyperparameters and the degrees-of-freedom parameter is discussed in the next subsection, where we also specify the covariance functions for the latent processes $f_1$ and $f_2$.

\subsection{Priors for the GP hyperparameters and degrees-of-freedom parameter}

When the parameter $\nu$ $\rightarrow$ $0$, the Student-$\it{t}$ model presents higher robustness, in which case the likelihood function may be unbounded and so difficult to evaluate (see \citealt{fernandes:1999} and \citealt{wang:2016}). Moreover, Gaussian process priors for the function values of the regression model introduce great flexibility into the model's fit capability. For which reason the model can perform poorly and present overfitted regression functions if the prior distributions are not carefully chosen for the covariance function hyperparameters \citep{simpson:2017}. 


With the goal of alleviating such scenarios, our choice in the prior distribution for the degrees-of-freedom $\nu$ follows the penalised model-component principles (PC), introduced by \cite{simpson:2017}. Under the hierarchical nature of the modelling approach, the main idea of PC-priors rest of the fact that the prior should avoid overly complex models whenever otherwise stated (see desideratas and principles in \cite{simpson:2017}). 

In this sense, we rather prefer a prior distribution for the degrees-of-freedom $\nu$ that does not favour too small values of $\nu$. Hence, we let $\nu \in (0, \infty)$ and, instead of imposing some kind of the restriction, e.g. $\nu$ $>$ $1$ \citep{van:2009, jyla:2011}, we choose a prior which does not favour values of $\nu$ $<$ $2$ (the variance \eqref{eq:predV} for the data does not exist in this case). Note that, it is the variance of a future outcome \eqref{eq:predV} that tells us about the uncertainty around the expected value \eqref{eq:predE} (point estimate). In all subsequent experiments, we will consider that $\nu$ $\sim$ $\mathrm{Gumbel\mbox{-}II}(1, \lambda)$, where $\lambda$ $=$ $- 2 \log \mathbb{P}(\nu < 2)$ and $\mathbb{P}(\nu < 2) = 0.1$. 

For the latent processes $f_1$ and $f_2$, we assume the squared exponential covariance function given by
\begin{equation} \label{eq:cov2}
\cov(f_j(\x), f_j(\x')|\sigma^2_j, \bell_j) = \sigma^2_j \exp \big(-\tfrac{1}{2}(\x - \x')^T [\diag(\bell_j)^2]^{-1} (\x - \x') \big)
\end{equation}
for $j$ $=$ $1$, $2$ and where the covariate space has dimension $p$, accordingly to each experiment in the next subsections. The vector of hyperparameters is given by $[\sigma^2_1 \ \bell_1 \ \sigma^2_2 \ \bell_2]$ where $\bell_1 = [\ell_{1, 1}, \cdots, \ell_{1, p}]^T$ and $\bell_2 = [\ell_{2, 1}, \cdots, \ell_{2, p}]^T$. The choice of the hyperpriors for the hyperparameters combines the wealkly informative principle from \cite{gelman:2006} and the PC-priors \citep{simpson:2017}. In this case, the density function for the hyperparameters should give more weight to simple regression functions (straight lines, planes, etc). That is it, the prior should favour small variability of the sample functions in the GP prior and more strongly correlated function values in order to avoid overfitting \citep[see][more for details]{gelman:2006, simpson:2017}. Hence, we assume that, $\sigma^2_1, \sigma^2_2$ $\stackrel{i.i.d}{\sim}$ $\mathcal{S}_{+}(0, \sigma^2_f, 4)$ for relatively small values of $\sigma_f^2$ and $\bell_1$, $\bell_2$ $\stackrel{i.i.d}{\sim}$ $\mathrm{inv}\mbox{-}\mathcal{S}_{+}(0, 1, 4)$. The specific choice for $\sigma_f^2$ will be given for each dataset in the subsequent sections. With this choice, the prior densities favour small variabilities of the Gaussian process prior for the function values and induce greater values of length-scales which increase the dependency between the function values. The notation $\mathrm{inv}\mbox{-}\mathcal{S}_{+}$ stands for inverse Student-$\it{t}$ distribution truncated on $\mathbb{R}_+$.


\subsection{Simulated data with simple regressions}

In this first experiment, we simulated a dataset tailored to work well with both approximate marginal likelihoods \eqref{eq:marglap} and \eqref{eq:marglapF}.
We then compared the Laplace approximations \eqref{eq:postap} and \eqref{eq:postap2} where we set $\btheta$ by either maximizing \eqref{eq:marglap} and \eqref{eq:marglapF} respectively. We consider that the probabilistic model for the data is given by \eqref{eq:stmodel} where $f_1(\cdot)$ and $f_2(\cdot)$ are unidimensional real-valued functions given by
\begin{align} 
f_1(x) &=  0.3 + 0.4x + 0.5\cos(2.7x) + \tfrac{1.1} {1 + x^2} \nonumber \\
f_2(x) &= 0.5 \cos(0.5 \pi x) + 0.52 \cos(\pi x) - 1.2.
\end{align}
Hence, the data generative mechanism is $Y|f_1(x), f_2(x), \nu \sim \mathcal{S}(f_1(x), \exp(f_2(x)), \nu)$ and the number of covariates is $p$ $=$ $1$. To simulate the dataset, we choose $\nu = 2.5$ and different sample sizes $n \in \lbrace 10, 150 \rbrace$ with equally spaced points in the interval $(-4.5, 4.5)$. The set of parameters $\btheta$ $=$ $[\nu$ $\sigma^2_1$ $\ell_1$ $\sigma^2_2$ $\ell_2]$ and we choose $\sigma_f^2$ $=$ $10$. The vector $\btheta$ (in the log scale) is either set by maximizing \eqref{eq:marglap}, which is denoted by $\btheta_1$, or by maximizing \eqref{eq:marglapF}, which is denoted by $\btheta_2$.

We compare the approximations \eqref{eq:postap} and \eqref{eq:postap2} by means of the estimated regression function $f_1(\cdot), f_2(\cdot)|\y, \btheta_r$ for $r = 1, 2$ and the local approximate posterior predictive distributions $f_1(x_*), f_2(x_*)|\y, \btheta_r$, at $x_* = 0$ for $r$ $=$ $1, 2$. 
\begin{table*}[!b]
\centering
\begin{tabular}{ll|lllllll}
  &      & $\nu $ &$\sigma_1^2$ & $\ell_1$ & $\sigma_2^2$ & $\ell_2$ & & \\ \hline
Maximum a posteriori $\btheta_1$ & $n = 10$  & 7.76  &  2.55  &    4.07  &  0.79   &  0.62 &  \\
 & $n = 150$ & 2.87  &  2.19  &    0.92  &  1.61   &  1.01 &    \\ \hline
Maximum a posteriori $\btheta_2$  & $n = 10$  & 4.66  &  2.52  &    3.52  &  0.38   &  0.98 &    \\
& $n = 150$ & 2.77  &  2.23  &    0.93  &  1.56   &  1.02 &   \\ \hline        
MCMC method   & $n = 10$  & 8.07  &  2.96  &    2.17  &  1.47   &  1.15 &   \\ 
(posterior mean)    & $n = 150$ & 2.74  &  2.44  &    0.90  &  1.39   &  1.02 &
 \end{tabular}
\caption{Maximum a posteriori estimates with different approximate marginal likelihoods and sample sizes. The estimate $\btheta_1$ corresponds to the value of $\btheta$ such that \eqref{eq:marglap} is maximized. The estimate $\btheta_2$ corresponds to the value of $\btheta$ such that \eqref{eq:marglapF} is maximized. 
The last row shows the posterior mean of $\btheta|\y$ estimated via MCMC approximation.}
\label{tab:tab_1}
\end{table*}
The natural gradient adaptation (equation \eqref{eq:amup}) is used to find $\hat{\f}$ for both approximations \eqref{eq:postap} and \eqref{eq:postap2}.
In both cases, the approximate marginal likelihoods \eqref{eq:marglap} and \eqref{eq:marglapF}
were stable to evaluate. Hence, $\btheta_1$ and $\btheta_2$ were obtained without any problems.
\begin{figure}[!htb]
\setlength{\parindent}{-1.45cm}
\includegraphics[scale = 0.38]{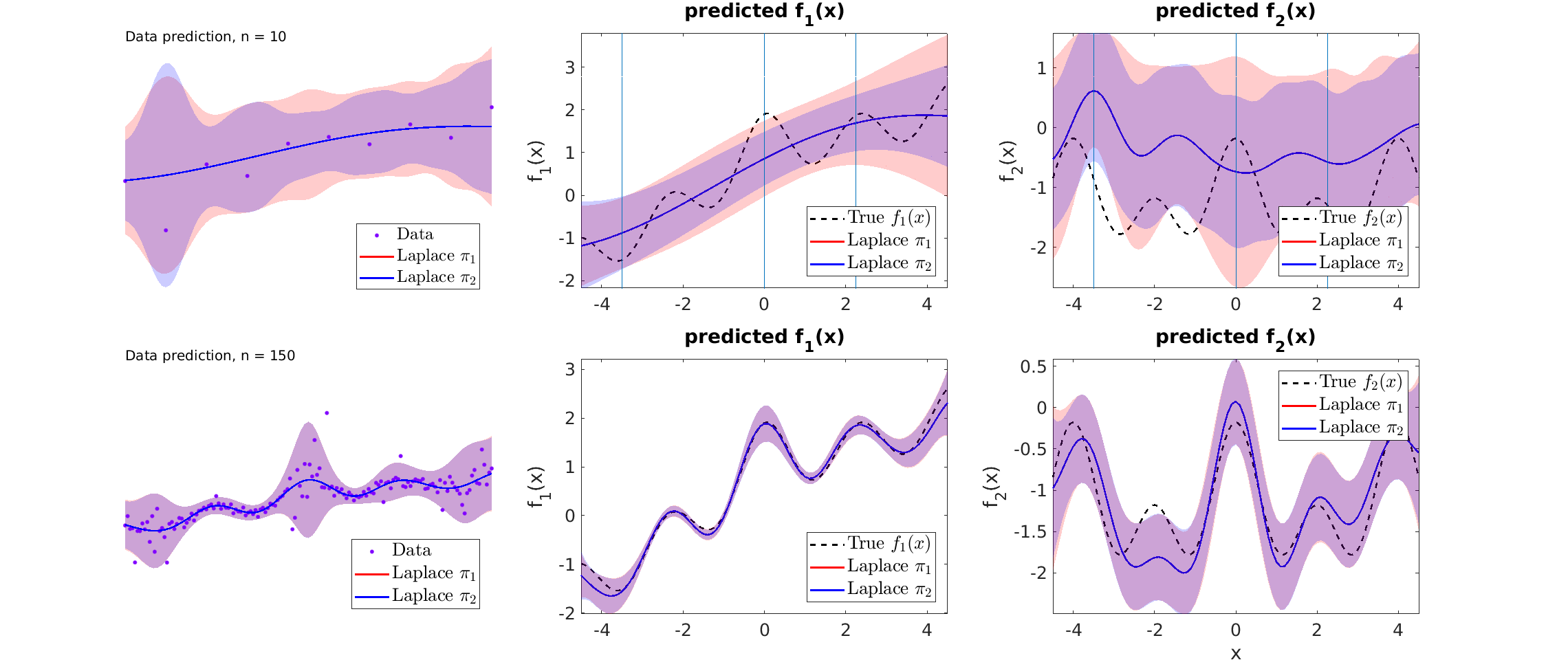}
\caption{Comparisons between the Laplace approximations \eqref{eq:postap} and \eqref{eq:postap2} where $\btheta = \btheta_1 $. In the first row, the sample size is $n = 10$ and in the second row the sample size is $n = 150$. The red color shows the approximate posterior predictive distributions for the regression functions $f_1(x)$ and $f_2(x)$ with the Laplace approximation \eqref{eq:postap}. The blue color shows the approximate posterior predictive distribution for the regression functions $f_1(x)$ and $f_2(x)$ with the Laplace-Fisher approximation \eqref{eq:postap2}. Note that, since $\btheta$ is the same in both Laplace approximations, the MAP estimate $\hat{\f}$ is the same for both approximations. In the second row, with a larger dataset, both approximations completely match.}
\label{fig:fig_4W}
\end{figure}

\begin{figure*}[t]
\setlength{\parindent}{-1.45cm}
\includegraphics[scale = 0.38]{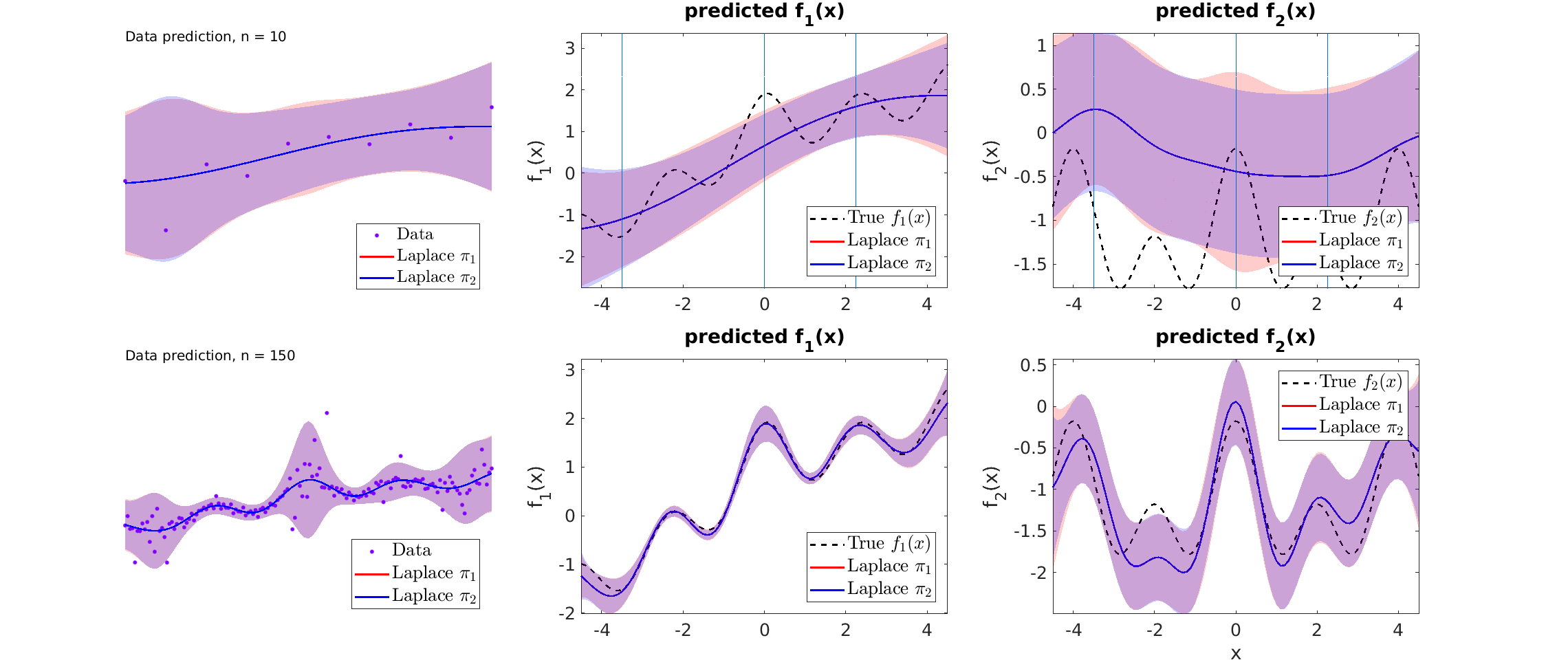}
\caption{Comparison between the Laplace approximations \eqref{eq:postap} and \eqref{eq:postap2} where $\btheta = \btheta_2 $. In the first row, the sample size is $n = 10$ and in the second row the sample size is $n = 150$. The red colour shows the approximate posterior predictive distribution for the regression functions $f_1(x)$ and $f_2(x)$ with the Laplace approximation \eqref{eq:postap}. The blue colour shows the approximate posterior predictive distribution for the regression functions $f_1(x)$ and $f_2(x)$ with the Laplace approximation \eqref{eq:postap2}. Note that, since $\btheta$ is the same in both Laplace approximations, the MAP estimate $\hat{\f}$ is the same for both approximations. In the first row the approximations are very similar and in the second row the approximations completely match each other again.}
\label{fig:fig_4F}
\end{figure*}

Table \ref{tab:tab_1} displays the maximum a posterior estimate for $\btheta$ using the approximate marginal likelihoods \eqref{eq:marglap} and \eqref{eq:marglapF}. The posterior mean of $\btheta|\y$ obtained with MCMC methods is also presented. Figure \ref{fig:fig_4W} and Figure \ref{fig:fig_4F} show the model performance for the Laplace approximations \eqref{eq:postap} and \eqref{eq:postap2} for $\btheta$ fixed as $\btheta_1$ and $\btheta_2$ respectively. In Figure \ref{fig:fig_4W}, the Laplace approximation \eqref{eq:postap} gives slighty different performance when compared to \eqref{eq:postap2} in the case where $n = 10$. In the case where $n = 150$, the approximations \eqref{eq:postap} and \eqref{eq:postap2} completely match. Figure \ref{fig:fig_4F} shows the result of the same experiment, however with $\btheta = \btheta_2$ for both approximations. We note that, for $n = 10$, the approximations \eqref{eq:postap} and \eqref{eq:postap2} show very similar performance. When $n = 150$, the approximations match again. In general, the Laplace approximations \eqref{eq:postap} and \eqref{eq:postap2} are slighty different for small sample sizes, but very similar when the number of data points increase, no matter whether $\btheta$ is chosen as $\btheta_1$ or $\btheta_2$.

\begin{figure}[!htb]
\setlength{\parindent}{0.6cm}
\includegraphics[scale = 0.43]{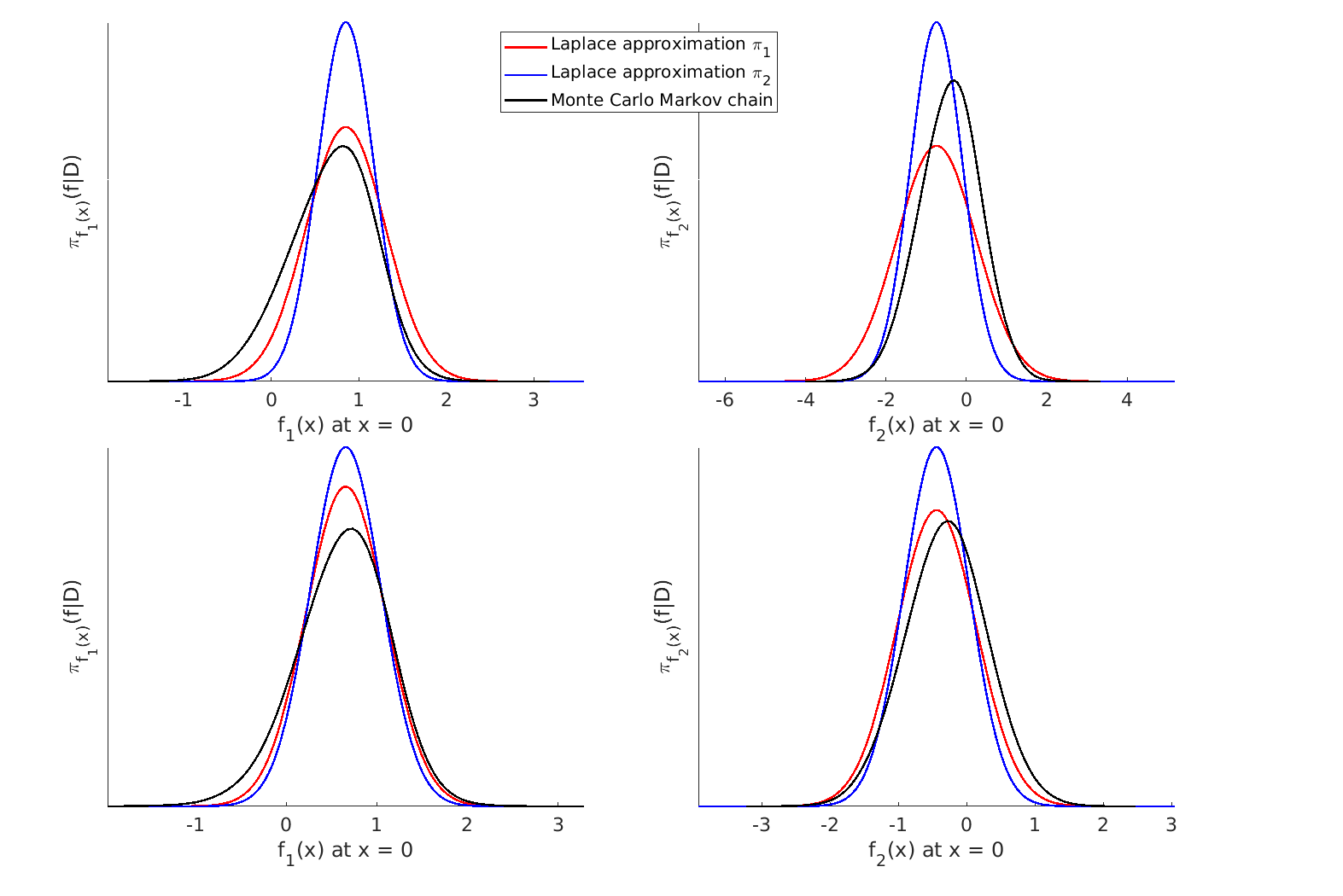}
\caption{Local comparisons between the approximate posterior predictive marginal distributions of the Laplace approximations \eqref{eq:postap}, \eqref{eq:postap2} and MCMC approximation. The upper row displays the approximate posterior predictive marginal distributions for $f_1(x)|\y, \btheta$ and $f_2(x)|\y, \btheta$ at $x = 0$ where $\btheta = \btheta_1$. The lower row displays the approximate posterior predictive marginal distribution for $f_1(x)|\y, \btheta$ and $f_2(x)|\y, \btheta$ at $x = 0$ where $\btheta = \btheta_2$. In all cases the dataset in the same and the sample size is $n = 10$.}
\label{fig:fig_3pw}
\end{figure}

In Figure \ref{fig:fig_3pw}, we compare the approximate posterior predictive distributions  \eqref{eq:postpred} with both Laplace approximations and with the MCMC approximation. We consider $x$ $=$ $0$ with the sample size $n = 10$. In the first row of Figure \ref{fig:fig_3pw}, all approximations of \eqref{eq:post} consider $\btheta$ $=$ $\btheta_1$. The Laplace-Fisher approximation estimates smaller variances in both cases. In the second row of Figure \ref{fig:fig_3pw}, we redo the same, but instead we set $\btheta = \btheta_2$. In this case, the difference between the approximate posterior predictive distributions whether considering the traditional Laplace approximation \eqref{eq:postap} or the Laplace-Fisher approximation \eqref{eq:postap2} is smaller than when $\btheta$ $=$ $\btheta_1$. The MCMC approximation for the true marginal predictive distribution also shows very similar performance.
\subsection{Predictive performance on real datasets}

In this section, the performance of the Laplace approximation \eqref{eq:postap} and \eqref{eq:postap2} for the Gaussian process regression with the heteroscedastic Student-$\it{t}$ model is examined with real data. Experiments with five datasets were conducted to evaluate the performance of different models in terms of predictive performance (see Appendix \ref{app:second} for a short description of the datasets).

We compare the predictive performance of the Laplace approximations \eqref{eq:postap} and \eqref{eq:postap2} with the Gaussian process regression with the homoscedastic Student-$\it{t}$ model \citep{van:2009} and the Gaussian process regression with the heteroscedastic Gaussian model. We also compare these models with the MCMC approximation of \eqref{eq:post} in the heteroscedastic Student-$\it{t}$ model. These models are respectively denoted by HT-ST-1, HT-ST-2, HM-ST, HT-G and HT-MCMC respectively.

The predictive preformance of the models were compared by splitting the datasets into training data ($n_{Training}$) and test data ($n_{Test}$), see Table \ref{tab:tab_2}. Three measures of predictive quality are proposed to compare all the models. 1) The absolute mean error $\mathcal{R}_1$ $=$ $1/n_{Test}$ $\sum_{i = 1}^{n_{Test}} |y_i - \mathbb{E}(Y_i|\btheta, \y)|$. 2) The root mean squared error $\mathcal{R}_2$ $=$ $(1/n_{Test} \sum_{i = 1}^{n_{Test}} (y_i - \mathbb{E}(Y_i|\btheta, \y))^2)^{\frac{1}{2}}$. 3) The log predictive density statistic $\mathcal{P}$ $=$ $\sum_{i = 1}^{n_{Test}} \log \pi(y_i|\btheta, \y)$ \citep[the greater the value of $\mathcal{P}$, the better the model is for the data analysis, see][]{gelman:2014}.



For all the models, inference on $\btheta$ is done by maximizing the respective approximate marginal likelihood \eqref{eq:marglap} and \eqref{eq:marglapF} of each Laplace approximation \eqref{eq:postap} and \eqref{eq:postap2} and $\hat{\f}$ is searched by the natural gradient method \eqref{eq:amup}. For model HM-ST, we set $\btheta$ by maximizing the approximate marginal likelihood as done by \cite{van:2009} and $\hat{\f}$ is obtained via the stabilized Newton algorithm \citep[see][page 3231, Section 3.2]{jyla:2011}. Model HT-G was implemented as HT-ST-1 with fixed $\nu = 5 \times 10^4$. In this case the Student-$\it{t}$ model practically corresponds to the Gaussian model.

Table \ref{tab:tab_2} shows the predictive performance for all the models with all the datasets. We see that all the models perform similarly with respect  $\mathcal{R}_1$ and $\mathcal{R}_2$. Model HT-G shows slighlty worse predictive performance with respect to $\mathcal{R}_1$, and this is reasonable. The Gaussian model for the data is not an outlier-prone model, if some training point $y_i$ is an outlier, then the predictive value $\mathbb{E}(Y_i|\btheta, \y)$ will try to match the training point. This is not the case with the Student-$\it{t}$ model for data. Note that, both of the statistics $\mathcal{R}_1$ and $\mathcal{R}_2$ use the discrepancy between $y_i$ and $\mathbb{E}(Y_i|\btheta, \y)$. In the case of $\mathcal{R}_2$, this discrepancy is squared, which penalizes the predictive quality of the model too much if the discrepancy for some particular data points are too high (or too small). With respect to the statistic $\mathcal{R}_1$, there is no harsh penalization. Hence the models HT-ST-1, HT-ST-2 shows slightly better predictive performance when compared to HT-GP. Overall, the model HT-S shows slightly better predictive performance with respect to $\mathcal{R}_1$ and $\mathcal{R}_2$, this means that this model tends to overfit to a small degree, since it does not allow for heteroscedasticity in the data.


Model HT-ST-2 has almost the same predictive performance as model HT-ST-1 with respect to $\mathcal{R}_1$ and $\mathcal{R}_2$. This is expected, given that the number of data points in all datasets are relatively high, the estimate of $\btheta$, whether from \eqref{eq:marglap} or \eqref{eq:marglapF} are similar. This implies similar $\hat{\f}$ in the approximations \eqref{eq:postap} and \eqref{eq:postap2}. Hence, according to equation \eqref{eq:predE}, the predictive measures $\mathcal{R}_1$ and $\mathcal{R}_2$ are close for both models HT-ST-1 and HT-ST-2. The performance of HM-ST has also shown good predictive performance with respect to $\mathcal{R}_1$ and $\mathcal{R}_2$ for all datasets, but it does not present good values with respect to statistic $\mathcal{P}$. Note, however, that $\mathcal{R}_1$ and $\mathcal{R}_2$ are measures of dispersion based on the estimate $\mathbb{E}(Y_i|\btheta, \y)$, which does not take into account the uncertainty in the predictive distribution of $Y_i|\y, \btheta$.
\begin{table}
\setlength{\parindent}{1.0cm}
  \begin{tabular}{llcccr}
     \toprule
  {\bf Dataset} & {\bf Models} & $\mathcal{R}_1$ &  $\mathcal{R}_2$  & $\mathcal{P}$ & {\bf Hyperpriors} \\ \hline
       Neal                   & HM-ST       & 0.08 & 0.14 & 84.50 & $\sigma^2_1, \sigma^2_2$ $\stackrel{i.i.d}{\sim}$ $\mathcal{S}_{+}(0, 15, 4)$ \\ 
       $n_{Training} = 100$   & HT-ST-1     & 0.08 & 0.14 & 84.80 & \\
       $n_{Test} = 100$       & HT-ST-2    & 0.08 & 0.14 & 85.13 & \\
       $p = 1$                & HT-G       & 0.09 & 0.14 & 53.51 & \\ 
                              & HT-MCMC & 0.08 & 0.13 & 85.13 & \\                            \hline
         Motorcycle           & HM-ST      & 20.06 & 26.78 & -316.38 & $\sigma^2_1, \sigma^2_2$ $\stackrel{i.i.d}{\sim}$ $\mathcal{S}_{+}(0, 500, 4)$ \\ 
         $n_{Training} = 67$ & HT-ST-1    & 20.50 & 27.31 & -297.44 &  \\
         $n_{Test} = 66$      & HT-ST-2    & 20.46 & 27.26 & -298.52 & \\
         $p = 1$             & HT-G       & 20.78 & 27.44 & -299.50 &  \\
                             & HT-MCMC    & 20.67 & 27.37 & -292.18 & \\                            \hline
         Boston                 & HM-ST      & 0.22 & 0.33 & -74.88 & $\sigma^2_1, \sigma^2_2$ $\stackrel{i.i.d}{\sim}$ $\mathcal{S}_{+}(0, 15, 4)$\\ 
        $n_{Training} = 253$  & HT-ST-1    & 0.25 & 0.41 & -51.49 & \\
        $n_{Test} = 253$      & HT-ST-2    & 0.25 & 0.41 & -52.64 & \\
       $p = 13$               & HT-G       & 0.26 & 0.41 & -60.08 & \\       
                              & HT-MCMC & 0.25 & 0.41 & -51.79 & \\ \hline
          Friedman             & HM-ST        & 1.56 & 1.98 & -214.61 & $\sigma^2_1, \sigma^2_2$ $\stackrel{i.i.d}{\sim}$ $\mathcal{S}_{+}(0, 15, 4)$ \\ 
          $n_{Training} = 100$ & HT-ST-1    & 1.28 & 1.72 & -192.83 &  \\
         $n_{Test} = 100$      & HT-ST-2    & 1.26 & 1.72 & -192.81 & \\
         $p = 5$               & HT-G  & 1.29 & 1.72 & -196.15 &  \\
                          & HT-MCMC & 1.25 & 1.69 & -189.08 & \\ \hline
          Compressive     & HM-ST      & 4.19  & 5.85 & -1593.36 & $\sigma^2_1, \sigma^2_2$ $\stackrel{i.i.d}{\sim}$ $\mathcal{S}_{+}(0, 500, 4)$ \\  
      $n_{Training} = 515$        & HT-ST-1  & 4.31  & 6.15 & -1591.00 &  \\
      $n_{Test} = 515$          & HT-ST-2    & 4.35  & 6.19 & -1593.33 & \\
      $p = 8$             & HT-G       & 4.32 & 6.10 & -1598.90 &  \\ 
						  & HT-MCMC  & 4.38 & 6.16 & -1569.60 & \\ \hline
   \bottomrule                        
  \end{tabular}
 \caption{Model comparisons. $\mathcal{R}_1$ stands for the absolute mean squared error, $\mathcal{R}_2$ is the root mean squared error and $\mathcal{P}$ is the log-predictive density statistics. The number $n_{Training}$ is the sample size, $n_{Test}$ is the number of test points and $p$ is the number of covariates for each dataset. The second column shows the models examined in the experiments and the last column shows the hyperpriors chosen for the Gaussian processes hyperparameters. The model abbreviations stand for: 1) HM-ST - Laplace approximation for the GP regression with the homoscedastic Student-$\it{t}$ model, 2) HT-ST-1 - Laplace approximation for the GP regression with the heteroscedastic Student-$\it{t}$ model, 3) HT-ST-2 - Laplace-Fisher approximation for the GP regression with the heteroscedastic Student-$\it{t}$ model, 4) HT-G - Laplace approximation for the GP regression with the heteroscedastic Gaussian model and 5) HT-MCMC - MCMC approximation for the GP regression with the heteroscedastic Student-$\it{t}$ model.}
\label{tab:tab_2}
\end{table}

With respect to the $\mathcal{P}$ statistics, model HT-ST-1 dominates when compared to the models HT-G and HM-ST. For the model HT-ST-2, the statistics $\mathcal{P}$ is only slightly smaller compared to HT-ST-1. These outcomes are still quite reasonable. The $\mathcal{P}$ statistics calculates the value of the predictive density for a future outcome at the measured values. If the random variable $Y_i|\y, \btheta$ has small variance, its predictive density does not cover much region of the sample space, therefore, if the mode of the predictive density function is distant from the observed value, the density $\pi(y_i|\y, \btheta)$ is small. On the other hand, if $Y_i|\y, \btheta$ has greater variance, its predictive density covers greater regions of the sample space, therefore, even if the mode is distant from the observation, the density function of $Y_i|\y, \btheta$ evaluated at $y_i$ will be higher. This is exactly what happens with the models HT-ST-1 and HT-ST-2. The predictive distributions of $Y_i|\y, \btheta$, with models HT-ST-1 and HT-ST-2 have similar expectations since, in both approximate posteriors \eqref{eq:postap} and \eqref{eq:postap2}, the estimates for $\hat{\f}$ are similar. However, since the approximate variance of $\f|\y, \btheta$ is generally higher in the approximation $\eqref{eq:postap}$, $\pi(y_i|\y, \btheta)$ will be wider (see equation \eqref{eq:predV}), hence leading to a higher $\mathcal{P}$ statistics. 

The aforementioned behaviour is also analogous to the case where the Gaussian model for the data is assumed, since the Gaussian density function will always have thinner tails compared to the Student-$\it{t}$ model. Once we have chosen the probabilistic approach to conduct the data analysis, the statistic $\mathcal{P}$ may be considered a better suitable measure of predictive quality since it takes into account the degrees of uncertainty which is encoded in the posterior predictive distributions \citep{bern:1994, Vehtari+Ojanen:2012}.

As expected, the HT-MCMC model presents very similar results with respect to the predictive measures $\mathcal{R}_1$ and $\mathcal{R}_2$ compared to all other models. This model also presents the best predictive performance with respect to the predictive measure $\mathcal{P}$. This is also confirmatory in the sense of the previous explanation about $\mathcal{P}$, since this model approximates the true predictive distributions $\pi(y_i|\btheta, \y)$ better than Laplace's method.

Even though model HT-ST-2 only had slighty worse predictive performance compared to HT-ST-1, model HT-ST-2 still provided very similar results in all predictive measures. This result suggests that the Laplace-Fisher approximation \eqref{eq:postap2}, based on the Fisher information matrix in place of the Hessian matrix of the negative log-likelihood function, can also be a good candidate to approximate the posterior density \eqref{eq:post}.

The optimization of \eqref{eq:post} based on the natural gradient also provided benefits compared to previous approaches. In our experiments, the natural gradient adaptation was always able to converge, whereas the Newton's method was very sensitive to initial values of $\f$ and to the values of the parameters $\btheta$  \citep[a general discussion on this is given by, e.g.,][]{van:2009,jyla:2011}. This is not unexpected. In the Newton update \eqref{eq:upW}, $(K^{-1} + \nW)^{-1}$ is not always positive-definite (as it should be in the traditional Newton's method) and if the initial value for $\f$ is far from the mode of \eqref{eq:post}, the Newton's method will not converge. 

In all the experiments with simulated and real datasets, the initial value for $\f_1$ $=$ $\0$  and for $\f_2$ $=$ $\boldsymbol{3}$ (a vector where each element is equal to 3). This choice means that $\sigma(\x)$ $=$ $\exp(3) \approx 20$, in other words, at initialization the data has "large" variance compared to the prior variance of $f_1$ everywhere in the covariate space. This also avoids possible multimodality of the posterior density \eqref{eq:postap} since the initial values for $\sigma(\x)$ are relatively high (see the analysis done by \citealt{van:2009} Section 3.4 and \citealt{jyla:2011} Section 5, second paragraph). This will help for many datasets, but for example, the motorcycle dataset, where the range of variation of that data goes from -130 to 100 (see \citealt{silverman:1985}, Figure 2), the initial value for $\f_2$ is far from optimal. However, we have not encountered any problem in optimization of \eqref{eq:postap} with any dataset using the natural gradient adaptation.

\section{Concluding remarks and discussion} \label{sec:7}

Recently, many approximative methods have been propose to approximate the posterior distribution of the Gaussian process model with homoscedastic Student-$\it{t}$ probabilistic model for the data \citep[see][]{van:2009, jyla:2011}. With a non log-concave likelihood, those methods require special treatment by tuning certain values in the mechanism of the estimation process to incur convergence in the computational algorithm (see \citealt{van:2009}, Section 4.2 and \citealt{jyla:2011}, Section 4). 

In this paper, we extended the models presented by \cite{van:2009} and \cite{jyla:2011}, by additionally modelling the scale parameter of the Student-$\it{t}$ model with a Gaussian process prior. In general, the Gaussian process regression with the heteroscedastic Student-$\it{t}$ model has been shown to perform very well. With respect to the statistic $\mathcal{P}$, it has shown the best performance when compared to known models such as the Gaussian process regression with the homocesdastic Student-$\it{t}$ model of \cite{van:2009} and the Gaussian process regression with the heteroscedastic Gaussian model for the data. 

\cite{saul:2016} introduced chained Gaussian processes, which uses variational methods to approximate the posterior distribution of the Gaussian process regression with the heteroscedastic Student-$\it{t}$ model for the data. Additionaly, their approach allow the use of large datasets via sparse GP approximations \citep{snelson:2005,titsias:09, hensman:2015}. Our methodology could easily be extended to include sparse GP approximation as well. However, in this work, we have focused in the aspects of parametrization in statitical models and exploited the orthogonal parametrization of the Student-$\it{t}$ model. Due to this particular property, we have recovered well-known algorithms \citep{Rasmussen+Williams:2006} to perform approximate inference with the Laplace approximation and with the Laplace-Fisher approximation.

Although the Laplace approximation based on the Fisher information matrix has already been proposed in the literature, its application in the context of Gaussian process regression has not been investigated yet. In our case, with the Student-$\it{t}$ model, this approximation delivered very similar results in the experiments with simulated and real datasets. Thus, the methodology presented here provides an alternative approximation method for Gaussian process regression. This also concerns approximation methods with other probabilistic models and parametrization in the same lines of \citet{kuss:2005} and \cite{nick+2008}. Moreover, the choice of the parameters $\btheta$ through the approximate marginal likelihood  $q_2(\y|\btheta)$ \eqref{eq:marglapF}, can also be seen as a new way of adapting the unknown covariance function hyperparameters and the probabilistic model parameters. In difficult cases, where the dataset leads to difficult evaluation of $q_1(\y|\btheta)$ \eqref{eq:marglap}, one can always use $q_2(\y|\btheta)$ to choose $\btheta$ and use the Laplace approximation $\pi_1(\f|\y, \btheta)$ \eqref{eq:postap} if wanted. 

We also point out that, there are two possible avenues of improvement in the optimization of \eqref{eq:postap} via natural gradient. Firstly, as studied by \cite{yang:1998}, \cite{amari:1998} and \cite{fuku:2000} the natural gradient adaptation is a robust learning rule in the sense that the method might avoid plateaus and local maxima. Hence, the natural gradient may be better suited than Newton's method given that \eqref{eq:post} is not guaranteed to be unimodal. Secondly, as empirically evaluated by \cite{honkela+raiko:2010}, the natural gradient might increase the convergence speed of the optimization method and there might be stability with the simplification of the computational code. The latter holds true. The structure of the natural gradient update \eqref{eq:amup} provides stable implementation. But it is hard to state whether the natural gradient will always provide faster convergence. Some theoretical studies of the convergence speed and statistical properties of the natural gradient can be found in \cite{martens:14}, Section 12.


By carefully noting the particular orthogonal parametrization of the Student-$\it{t}$ model, the natural gradient for finding the parameters of the Laplace approximation proposed here becomes attractive. With this approach the Laplace approximation is available for non-log-concave likelihoods and likelihoods that depend on more than one Gaussian process with the same stability and easiness of implementation as the Laplace approximation for log-concave likelihoods presented by \cite{Rasmussen+Williams:2006} (see their book for pseudocode). 

The choice of the matrix of metric coefficient $G$, which may be difficult to obtain in general optimization settings, can always be induced through the probabilistic model for the data. Thus, due to the probabilistic nature of our approach, the natural gradient is better suited to optimize the posterior density of the Gaussian process than the Newton's method. Moreover, for the most of the probabilistic models presented in the literature, the Fisher information matrix is available in closed-form \citep[see][]{john:1995}. Hence, one can always investigate a new parametrization for the probabilistic model such that the Fisher information matrix is diagonal (see Section 2). Besides, this is not restricted to the case where two parameters of a probabilistic model are modelled with Gaussian process priors, as shown in this paper. In fact, the approach presented here can also be used in the homoscedastic Student-$\it{t}$ model of \cite{van:2009} as well as in other uniparametric models, such as the Bernoulli and Poisson. These uniparametric models are commonly used within the context of Gaussian process regression and some type of reparametrization could be beneficial to improve posterior approximations and the estimation process. The studies by \cite{achcar:1990}, \cite{kass:1994}, \cite{achcar:1994} and \cite{mackay:1998} indicate and discuss possible ways to do so.

More generally, concepts of reparametrization in statistical modelling within the Gaussian process regression context deserve more attention. There is freedom of choice in the parametrization of the probabilistic model. If the posterior "normality" or inferential procedures can be improved under a different parametrizations, then approximation methods may be reassessed. That is, all of the well known approximation methods such as variational-Bayes, expectation-propagation or Laplace's method, approximate the target density with a Gaussian density. If the target density in some new parametrization is closer to a Gaussian, then the choice of the approximation method may not be as crucial as its computational aspects.

These aspect of reparametrization are also important for MCMC methods. If there are difficulties to sample from a posterior density in some specific parametrization of the model, one can also investigate a new parametrization so that the sampling problem is alleviated. For example, in the state-of-the-art Riemann manifold Hamiltonian Monte Carlo method (RMHMC) \citep{benmark11} the choice of the Riemannian metric (the Fisher information matrix) is essential for achieving good performance of the sampler. However, its computational implementation is hard and costly since $G$ is full matrix in most practical applications. If there is a possibility to find an orthogonal parametrization for the model parameters such that $G$ is diagonal, or at least it is not full matrix, then the computational aspects of the method could be further simplified. In this sense, the attractiveness of the method due to its properties would increase its use in practical applications.

The code implementing the model and the natural gradient approach as well as the Newton method are freely available at [link to be provided after acceptance]. A demo code also follows in the aforementioned link.

\section*{Acknowledgements}
The work was funded by the Academy of Finland (Grant 304531) and the research funds of University of Helsinki. The first author thanks Luiz Roberto Hartmann Junior, Susan Chumbimune, Marco Pollo Almeida, Teodoro Calvo and Donald Smart for their comments and suggestions that helped improve the paper.

\newpage

\appendix

\section{Datasets} \label{app:second}

A short description of the benchmark datasets used to evaluate the predictive performance of the models proposed in this paper. See Section \ref{sec:6}, Table \ref{tab:tab_2}. \\

\noindent
{\bf Neal}. This is a simulated dataset with the presence of strong outliers. The dataset was also used by \cite{neal:1997} (see page 21, Figure 5) for the Gaussian process regression with the homocesdastic Student-$\it{t}$ model. 

\noindent
{\bf Motorcycle}. This dataset consists of motorcycle accelerometer readings versus the time of impact in order to study the efficacy of helmets. This case ilustrates a unidimensional nonlinear regression problem which was studied by \cite{silverman:1985}. 

\noindent
{\bf Boston housing}. A well-known study case on housing prices, which was used to investigate whether clean air influenced the price of houses within the Boston metropolitan area in 1978. The dataset is composed by 506 measurements (census tracts) where each measurement consists of 13 covariates and 1 dependent variable, which is the median house price for that tract. The detailed description of each explanatory variable can be consulted in \cite{harretal:1978} table IV. 


\noindent
{\bf Friedman}. A special regression function provided by \cite{friedman:1991} and \cite{jyla:2011}, which involves a nonlinear regression function with 5 covariates. To make the experiment more challenging for the inference algorithm, 5 extra random covariates were generated as described by \cite{jyla:2011}. In this experiment a dataset with 200 observations is generated with 10 randomly selected outliers. 

\noindent
{\bf Compressive}. A dataset for which the task is to predict concrete compressive strength based on 8 covariates and 1030 measurements. More details are described in \cite{yeh:1998}. 

\section{Extra formulas} \label{app:thirdd}

In all the equations presented below we consider that $
z_i$ $=$ $\frac{y_i - f_1(\x_i)}{\exp(f_2(\x_i))}$ for $i$ $=$ $1,\hdots,n$.

\subsection{The elements of the matrix $\nW$}

\begin{equation}
\nW_{i, j} = \left\{
\begin{array}{l} \tfrac{1}{[\exp(f_2(\x_i))]^2} \big(1 + \tfrac{1}{\nu} \big) \left[\tfrac{2}{(1 + z^2_i/\nu)^2 } - \tfrac{1}{1 + z_i/\nu} \right], \ \mathrm{for} \  i = j = 1, \ldots, n
\\\\
\tfrac{2}{\exp(f_2(\x_i))}  \big(1 + \tfrac{1}{\nu} \big) \tfrac{z_i}{(1 + z^2_i/\nu)^2}, \\[0.4cm] \mathrm{for} \ i = 1, \ldots, N \ \mathrm{and} \ j = (i + n)\mathds{1}_{\lbrace 1, \hdots,n\rbrace}(i) + (i - n)\mathds{1}_{\lbrace n+1, \hdots, N\rbrace}(i)  \\\\
-1 + \big(1 + \tfrac{1}{\nu} \big)\tfrac{z_i^2}{(1 + z^2_i/\nu)}\Big[1 + \tfrac{2}{(1 + z^2_i/\nu)}\Big] + \tfrac{z_i^2 -1}{(1 + z^2_i/\nu)}, \ \mathrm{for} \ i = j = n+1, \ldots, N 
\\\\
0, \ \mathrm{otherwise}.
\end{array} 
\right.
\end{equation}

\subsection{The elements of the Fisher information matrix $\mathbb{E}_{\Y|\f, \btheta}[\nW]$}
\begin{equation} \label{eq:fisher}
\mathbb{E}_{\Y|\f, \btheta}[\nW]_{i, j} = \left\{
\begin{array}{l} \tfrac{\nu + 1}{\nu + 3}\exp(-2f_2(\x_i)), \ \mathrm{for} \  i = j = 1, \ldots, n
\\\\
\frac{2\nu}{\nu+3}, \ \mathrm{for} \ i = j = n+1, \ldots, N
\\\\
0, \ \mathrm{otherwise}.
\end{array} 
\right.
\end{equation}

\subsection{Derivatives of the $\log L(\y|\f, \nu)$ and {\normalfont W}} 

For each $i = 1, \ldots,n$ the elements of the gradient $\nabla_{\f} \log L(\y|\f, \nu)$ are given by
\begin{align}
\tfrac{\partial \log \pi(y_i|f_1(\x_i), f_2(\x_i), \nu)}{\partial {f_1(\x_i)}} =&  \big(1 + \tfrac{1}{\nu} \big) \tfrac{z_i}{\exp(f_2(\x_i))(1 + z^2_i/\nu)} \nonumber \\
\tfrac{\partial \log \pi(y_i|f_1(\x_i), f_2(\x_i), \nu)}{\partial {f_2(\x_i)}} =& \tfrac{z_i^2 - 1}{(1 + z^2_i/\nu)} 
\end{align}
and
\begin{equation}
\tfrac{\partial \log L(\y|\f, \nu)}{\partial \nu} =  \tfrac{n}{2} \psi\big(\tfrac{\nu + 1}{2}\big) - \tfrac{n}{2} \psi\big(\tfrac{\nu}{2}\big) - \tfrac{n}{2\nu} - \sum_{i = 1}^n \log\big(1 + \tfrac{1}{\nu}z_i^2\big) + \tfrac{z_i^2 (\nu + 1)}{\nu^2 (1 + z^2_i/\nu)} 
\end{equation}
The elements of the derivatives of $\nabla_{\f} \log L(\y|\f, \nu)$ w.r.t $\nu$ are
\begin{align}
\tfrac{\partial^2\log \pi(y_i|f_1(\x_i), f_2(\x_i), \sigma, \nu)}{\partial \nu \partial {f_1(\x_i)}} =& \tfrac{2}{[\exp(f_2(\x_i))]} \tfrac{z_i^3-z}{\nu^2(1 + z^2_i/\nu)^2}  \nonumber \\
\tfrac{\partial^2 \log \pi(y_i|f_1(\x_i), f_2(\x_i), \sigma, \nu)}{\partial \nu \partial f_2(\x_i)} =& \tfrac{z_i^4-z_i^2}{\nu^2(1 + z^2_i/\nu)^2}
\end{align}
The derivatives of each element of $\nW$ w.r.t $\f_1$, $\f_2$ and $\nu$ are given subsequently. Note that these are third-order derivatives of the negative of the log-likelihood function and so some derivatives will appear twice since the order of the derivatives can be interchanged. 
\begin{equation}
\tfrac{\partial \hspace{-0.03cm} \nW}{\partial f_1(\x_i)} = \left\{
\begin{array}{l} \tfrac{1}{[\exp(f_2(\x_i))]^3} \big(1 + \tfrac{1}{\nu} \big) \left[\tfrac{2z_i}{\nu(1 + z^2_i/\nu)^2 } \big(\tfrac{4}{1+z_i^2/\nu} - 1\big) \right], \ \mathrm{for} \ i = j = 1, \ldots, n
\\\\ 

-\tfrac{2}{[\exp(f_2(\x_i))]^2} \big(1 + \tfrac{1}{\nu} \big) \tfrac{1}{(1 + z^2_i/\nu)^2}\left[1 - \tfrac{4z^2_i}{\nu(1 + z^2_i/\nu)} \right], \\[0.4cm] \mathrm{for} \ i = 1, \ldots, N \ \mathrm{and} \  j = (i + n)\mathds{1}_{\lbrace 1, \hdots,n\rbrace}(i) + (i - n)\mathds{1}_{\lbrace n+1, \hdots, N \rbrace}(i) \\\\

- \tfrac{1}{\exp(f_2(\x_i))} \big(1 + \tfrac{1}{\nu} \big) \tfrac{z_i}{(1 + z^2_i/\nu)^2} \left[4 - \tfrac{8z^2_i}{\nu(1 + z^2_i/\nu)} \right], \ \mathrm{for} \ i = j = n+1, \ldots, N
\\\\
0, \ \mathrm{otherwise}.
\end{array} 
\right.
\end{equation}
%
%
\begin{equation}
\tfrac{\partial \hspace{-0.03cm} \nW}{\partial f_2(\x_i)} = \left\{
\begin{array}{l} 

-\tfrac{2}{[\exp(f_2(\x_i))]^2} \big(1 + \tfrac{1}{\nu} \big) \tfrac{1}{(1 + z^2_i/\nu)^2}\left[1 - \tfrac{4z^2_i}{\nu(1 + z^2_i/\nu)} \right], \ \mathrm{for} \ i = j = 1, \ldots, n
\\\\ 

- \tfrac{1}{\exp(f_2(\x_i))} \big(1 + \tfrac{1}{\nu} \big) \tfrac{z_i}{(1 + z^2_i/\nu)^2} \left[4 - \tfrac{8z^2_i}{\nu(1 + z^2_i/\nu)} \right], \\[0.4cm]\ \mathrm{for} \ i = 1, \ldots, N \ \mathrm{and} \ j = (i + n)\mathds{1}_{\lbrace 1, \hdots, n\rbrace}(i) + (i - n)\mathds{1}_{\lbrace n+1, \hdots, N\rbrace}(i) \\\\

 2 - \big(2 + \tfrac{2}{\nu} \big)\left[2z_i^2 + \tfrac{4z_i^2 - z_i^4/\nu}{(1 + z^2_i/\nu)} - \tfrac{4z^4_i}{\nu(1 + z^2_i/\nu)^2} \right] - \tfrac{z^2_i - 1}{(1 + z^2_i/\nu)} + \hdots \\[0.4cm] 3 \left[ -1 + \big(1 + \tfrac{1}{\nu} \big) \tfrac{z^2_i}{(1 + z^2_i/\nu)} \big(1 + \tfrac{2}{(1 + z^2_i/\nu)} \big) \right] , \\[0.4cm] \mathrm{for} \ i = j = n+1, \ldots, N 
\\\\
0, \ \mathrm{otherwise}.
\end{array} 
\right.
\end{equation}
\begin{equation}
\tfrac{\partial \hspace{-0.03cm} \nW}{\partial \nu} = \left\{
\begin{array}{l} 

\tfrac{1}{\exp(f_2(\x_i))]^2} \Big\{ -\tfrac{1}{\nu^2} \left[ \tfrac{2}{(1 + z^2_i/\nu)^2} - \tfrac{1}{(1 + z^2_i/\nu)} \right] + \big(1 + \tfrac{1}{\nu} \big) \left[ \tfrac{4z^2_i}{(1 + z^2_i/\nu)^2} - \tfrac{z^2}{\nu(1 + z^2_i/\nu)} \right] \Big\}
\\[0.4cm] \mathrm{for} \ i = j = 1, \ldots, n \\\\ 

-\tfrac{2}{\exp(f_2(\x_i))]} \tfrac{1}{(1 + z^2_i/\nu)^2} \left[\tfrac{z}{\nu^2} - \big(1 + \tfrac{1}{\nu} \big) \tfrac{2z^3_i}{\nu^2(1 + z^2_i/\nu)^2} \right], \\[0.4cm] \ \mathrm{for} \ i = 1, \ldots, N \ \mathrm{and} \ j = (i + n)\mathds{1}_{\lbrace 1, \hdots,n\rbrace}(i) + (i - n)\mathds{1}_{\lbrace n+1, \hdots,N\rbrace}(i) \\\\

-\tfrac{2z_i^2}{\nu^2} \left[\tfrac{2}{(1 + z^2_i/\nu)^2} + \tfrac{1}{(1 + z^2_i/\nu)} \right] + \big(1 + \tfrac{1}{\nu} \big) \tfrac{z_i^2}{\nu^2}\left[\tfrac{4}{(1 + z^2_i/\nu)^3} + \tfrac{1}{(1 + z^2_i/\nu)^2} \right] \hdots \\[0.4cm] 
+ \tfrac{z_i^4 - z_i^2}{\nu^2(1 + z^2_i/\nu)^2}, \ \mathrm{for} \ i = j = n+1, \ldots, N 
\\\\
0, \ \mathrm{otherwise}.
\end{array} 
\right.
\end{equation}

\vskip 0.2in

\bibliographystyle{apa-good}
\bibliography{refs}

\end{document}